\documentclass[twocolumn,a4paper,10pt]{article}
\usepackage{amsmath}
\usepackage[nointegrals]{wasysym}
\usepackage[T1]{fontenc}
\usepackage{esint}
\usepackage[total={6.7in,9.2in}]{geometry}
\usepackage{authblk}
\usepackage{hyperref}
\usepackage{abstract}
\usepackage[symbol]{footmisc}
\usepackage{placeins}
\usepackage[title,page]{appendix}
\usepackage{amssymb}

\usepackage{scalerel}
\usepackage{stackengine,wasysym}

\usepackage{nicefrac}

\usepackage{amsfonts}
\usepackage{makecell}
\usepackage{xcolor}
\usepackage{bbm}
\usepackage{graphicx}
\usepackage{bm}

\usepackage{cleveref}
\Crefname{equation}{Eq.}{Eqs.}
\Crefname{figure}{Fig.}{Figs.}
\Crefname{appendix}{SI}{SI}

\newcommand{\mailto}[1]{\href{mailto:#1}{#1}}

\usepackage[version=4]{mhchem}
\usepackage{bm}

\newcommand{\E}{\mathbb{E}}

\newcommand{\dif}{\mathrm{d}}

\def\eps{\varepsilon}

\DeclareMathOperator{\Geom}{Geom}

\newcommand{\given}{{\, | \,}}

\DeclareMathOperator{\Var}{Var}

\usepackage[labelfont=bf]{caption}
\usepackage{subcaption}

\usepackage[sorting=none,
            style=numeric-comp,
            bibstyle=phys,
            url=false,
            doi=false,
            isbn=false,
            language=english,
            clearlang=true,
            maxnames=10,
            eprint=true,
            maxbibnames=10,
            giveninits=true]{biblatex}

\AtEveryBibitem{\clearfield{month}}
\AtEveryBibitem{\clearfield{day}}
\AtEveryBibitem{\clearfield{note}}
\AtEveryBibitem{\clearfield{url}}
\AtEveryBibitem{\clearfield{doi}}

\AtEveryBibitem{\clearlist{language}}

\AtEveryBibitem{
 \ifentrytype{article}{
  \clearfield{eprint}
  \clearfield{eprinttype}
  \clearname{editor}
 }{}
}

\AtEveryBibitem{
 \ifentrytype{book}{
  \clearfield{series}
  \clearfield{volume}
  \clearfield{location}
 }{}
}

\AtEveryBibitem{
  \clearfield{eprintclass}}
  
\newcommand{\spr}{{\rho}}

\newcommand{\LRA}{{{}\;\;\Leftrightarrow\;\;{}}}

\title{\vspace{-1cm} The two-clock problem in population dynamics}

\author[ ]{Kaan \"Ocal$^*$}
\author[ ]{Michael P.~H.~Stumpf}
\affil[ ]{\centering \normalsize School of BioSciences \& School of Mathematics and Statistics \protect\\ University of Melbourne \protect\\ Parkville, Victoria 3052, Australia}
\date{}

\begin{document}

\twocolumn[
\maketitle 

\centering
\begin{minipage}{0.8\linewidth}

\centering 
\vspace{-0.5cm}
\small 
\begin{minipage}{0.95\linewidth}
\begin{abstract}
\noindent Biological time can be measured in two ways: in generations and in physical (chronological) time. When generations overlap, these two notions diverge, which impedes our ability to relate mathematical models to real populations. In this paper we show that nevertheless, the two clocks can be synchronised in the long run via a simple identity relating generational and physical time. This equivalence allows us to directly translate statements from the generational picture to the physical picture and vice versa. We derive a generalized Euler-Lotka equation linking the basic reproduction number $R_0$ to the growth rate, and present a simple identity that relates the selection coefficient of a mutation to the history of typical individuals, with applications to epidemiology, population biology and microbial growth.
\end{abstract}

\end{minipage}
\end{minipage}

\bigbreak

{}
]


\footnotetext[1]{\mailto{kaan.ocal@unimelb.edu.au}}

\section*{Introduction}

The asynchronous nature of reproduction is a fundamental difficulty when describing biological populations. We typically understand populations in terms of discrete generations, but physical populations consist of individuals from many different generations at once. Biological populations are thus simultaneously governed by two different clocks, one measuring generations, the other physical (or chronological) time. Since generation lengths can often differ between individuals, these two clocks desynchronise over time as some individuals procreate faster than others. This poses an inconvenience for population modeling, which frequently requires us to switch from the generational to the physical clock and vice versa depending on the application. At the heart of population biology therefore lies the question of how we can reconcile the generational and physical views of a population \cite{jagers_general_1989}.

\begin{figure*}[t]
    \centering
    \includegraphics{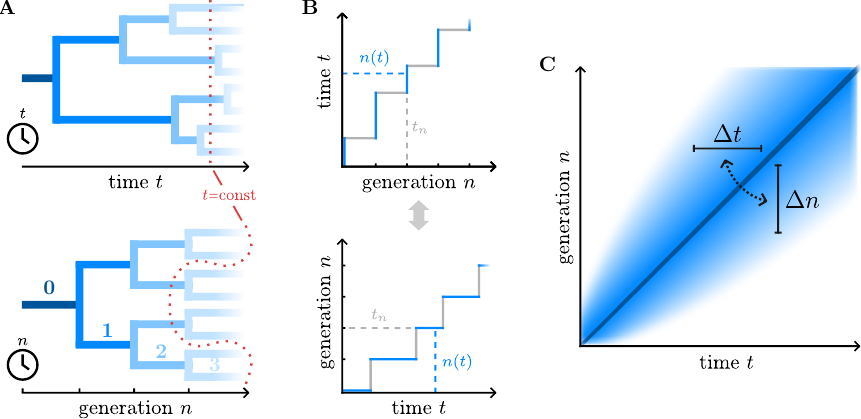}
    \caption{Population growth in the $n$- and $t$-picture. \textbf{A} Two views of a growing population. The $t$-ensemble (top) consists of individuals present at a fixed physical time $t$ (dashed line), whereas the $n$-ensemble (bottom) considers all individuals in a fixed generation. When generation lengths vary, the two ensembles are different: one cannot map a physical time $t$ to a unique generation. \textbf{B} In the $n$-ensemble a lineage is defined by the time $t_n$ it takes to reach generation $n$. In the $t$-ensemble, the same lineage is defined by its generation $n(t)$ at time $t$. These two viewpoints are mathematically equivalent. \textbf{C} Fluctuations $\Delta t_n$ in the $n$-ensemble directly correspond to fluctuations $\Delta n(t)$ in the $t$-ensemble. In the long-time limit, this correspondence has a very simple description in terms of large deviation theory.}
    \label{fig:intro}
\end{figure*}

We formalize this disconnect by distinguishing the $n$-picture of a population, which describes it in terms of generations, and the $t$-picture, which uses physical time (see Fig.~\ref{fig:intro}A). Population models track individuals proliferating across generations in the $n$-picture, whereas our physical experiences follow the $t$-picture. To model biological populations we often have to translate from the $n$-picture to the $t$-picture and back, but apart from a collection of techniques and heuristics that perform said translation, little is known about the general relationship between the two. Since differences in generation lengths can accumulate over time, the two clocks will progressively drift apart, which suggests that obtaining consistent predictions in both two pictures might be fundamentally difficult. Perhaps surprisingly, we show that this is not so: the two clocks are \emph{asymptotically equivalent}.

The problem of aligning the two clocks dates back to early studies of population growth due to Euler and Lotka \cite{lotka_relation_1907}, who considered the relationship between the basic reproductive number $R_0$ (the average offspring per individual) and the physical growth rate, or Malthus parameter $\Lambda$. These determine the population size $N$ in generation $n$ and at time $t$ via the asymptotics
\begin{align}
    N_n &\sim R_0^n, & N(t) &\sim e^{\Lambda t}.  \label{eq:first}
\end{align}

\noindent Thus $R_0$ is the growth rate in the $n$-picture, corresponding to $\Lambda$ in the $t$-picture. A fundamental principle of population biology \cite{heesterbeek_concept_1996} states that
\begin{align}
    \Lambda > 0 &\LRA R_0 > 1, \label{eq:epi_fundamental}
\end{align}

\noindent which characterises growing (as opposed to shrinking) populations, but beyond this, the relationship between the two growth rates is not well understood. For instance, models of cell division often assume that $R_0 = 2$, so that microbial fitness is exclusively determined by differences in division times \cite{nozoe_inferring_2017,hashimoto_noise-driven_2016,lin_single-cell_2020,genthon_noisy_2025}. On the other hand, in epidemics one can usually estimate the physical growth rate $\Lambda$ from data, but the threshold for herd immunity is determined by $R_0$, the inference of which is highly model-dependent \cite{heffernan_perspectives_2005,wallinga_how_2006}. 

For simple population models, the Euler-Lotka equation \cite{lotka_relation_1907} establishes a relationship between the two growth rates: if each individual in the population has a random lifetime $\tau$ independently sampled from a distribution $f(\tau)$ and produces an average of $R_0$ offspring upon death, then
\begin{align}
    R_0 \int_0^\infty f(\tau) \, e^{-\Lambda \tau} \dif \tau &= 1. \label{eq:el}
\end{align}

\noindent This equation and generalizations thereof \cite{powell_growth_1956,lebowitz_theory_1974,pigolotti_generalized_2021} have hitherto formed the basis for our understanding of population growth in real time, connecting the $n$-picture with the $t$-picture. In this paper, we provide a unifying framework for the two pictures that generalizes the Euler-Lotka equation and allows us to convert questions posed in one picture (such as computing growth rates and selection coefficients) into the other. 

Our framework is based on ideas from thermodynamics and statistical physics and proceeds by analyzing the fluctuations in generation lengths across lineages. Differences in generation lengths accumulate over time and these cannot be neglected, even in the long run: they fundamentally shape the behavior of a population. Nevertheless, the large-scale patterns exhibited by these fluctuations eventually give rise to predictable behavior. This is similar to thermodynamical systems such as an ideal gas, where the random movement of individual molecules, which is unpredictable on a microscopic level, is responsible for macroscopic phenomena such as pressure and temperature. We capitalize on this analogy and describe the $n$-picture and the $t$-picture in terms of two thermodynamic ensembles that turn out to be mathematically equivalent. This equivalence allows us to derive a systematic way to convert questions from one picture into the other, and formally resembles the well-known equivalence between the microcanonical and canonical ensembles in statistical mechanics. 

Our work directly builds upon recent results in \cite{ wakamoto_optimal_2012,levien_large_2020,pigolotti_generalized_2021}, which use large deviation theory to characterize the long-term behaviour of lineages. Understanding populations by analyzing the behavior of individual lineages has become a very fruitful avenue in population biology \cite{jagers_stabilities_1992,georgii_supercritical_2003,wakamoto_optimal_2012,nozoe_inferring_2017}, and is reminiscent of the path-integral formalisms used in other branches of physics, such as statistical mechanics and quantum field theory. Thermodynamical approaches that view a population as a macroscopic entity consisting of many interacting lineages have proven particularly popular \cite{garcia-garcia_linking_2019,genthon_fluctuation_2020,levien_large_2020}, and are closely related to large deviation theory \cite{touchette_large_2009}. 

A fundamental observation about populations is that forward lineages do not describe the ``average'' individual in a population. Since not all individuals are equally fit, the distribution of ancestral lineages, which characterize the history of a typical organism, will be skewed towards more successful individuals \cite{georgii_supercritical_2003,wakamoto_optimal_2012,nozoe_inferring_2017}. Understanding how selection shapes the histories and genealogies of individuals is necessary to model dynamical processes taking place in a population, such as inheritance, mutation and gene expression \cite{georgii_supercritical_2003,leibler_individual_2010,thomas_making_2017,thomas_intrinsic_2019,stadler_phylodynamics_2021,ocal_cell_2025}. We therefore describe ancestral, or backward lineages from a general thermodynamic viewpoint and show how they can be defined consistently in both pictures. The statistical properties of backward lineages encode many fundamental properties of a population, in particular how it responds to change. To this end, we obtain a very general formula for the selection coefficient of a mutation based on backward lineages, directly connecting our work with classical population genetics. We furthermore derive several Jensen-like inequalities that relate $R_0$, $\Lambda$, and the statistics of individual lineages, extending previous work in \cite{garcia-garcia_linking_2019,yamauchi_unified_2022}. 

Our paper follows in a line of recent work \cite{gingrich_fundamental_2017,raghu_thermodynamic_2025,duffy_how_2005} establishing asymptotic relationships fluctuations in thermodynamic quantities at a fixed time (the $t$-ensemble) with fluctuations in first passage times of these currents (the $n$-ensemble). While the distinction between the two ensembles is usually not made explicit, recent work has shown that a careful treatment of the two can lead to new techniques and insights in analysing population processes \cite{grandpre_extremal_2025,cure_exponential_2025}.

%


\section*{Theory}


\subsection*{Lineages and populations}

\label{sec:pop}

\begin{figure*}
    \centering
    \includegraphics{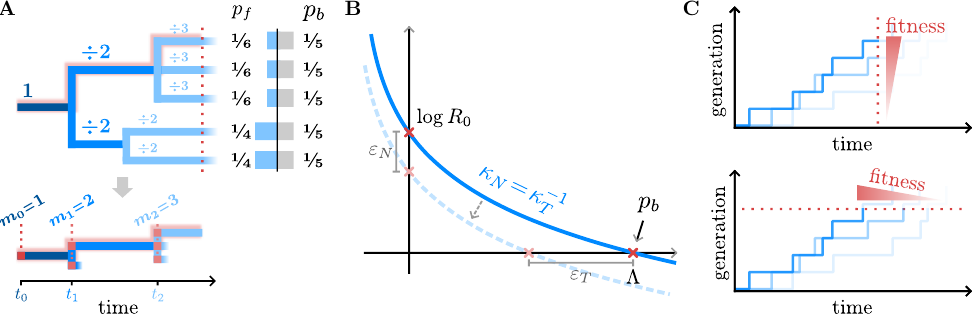}

    \caption{Statistical properties of lineages. \textbf{A} The forward distribution over lineages in a population starts with the original ancestor and follows one offspring at random in each generation. The forward probability of a lineage depends on its multiplicities $m_k$, and is lower for individuals with more siblings. In contrast, the backward distribution weighs each individual in the current population equally. \textbf{B} The log-partition functions $\kappa_N$ and $\kappa_T$ determine how changing $\log R_0$ by $\eps_N$ in the $n$-picture affects the growth rate $\Lambda$ in the $t$-picture, and vice versa. Its behavior around the point $\kappa_T(0) = \Lambda$ encodes information about the backward distribution $p_b$ over lineages, and therefore the statistics of the population as a whole. \textbf{C} The fitness of a lineage in the $t$-picture is measured by its reproductive success up to time $t$, which depends on its absolute fitness $w(t)$. In the $n$-picture, the fitness of a lineage is defined by its absolute fitness $w_n$ and the time $t_n$ it takes to reach generation $n$. Asymptotically, the two notions of fitness coincide.}
    \label{fig:lineages}
\end{figure*}

For our purposes, a population consists of all lineages that stem from one common ancestor, visualised as a tree in Fig.~\ref{fig:lineages}A. A lineage is a sequence of individuals $x_0, x_1, \ldots$ that are direct descendants of each other, starting with the common ancestor $x_0$. The $i$-th individual is born at time $t_i$ and has $m_i$ siblings including itself, where by definition $t_0 = 0$ and $m_0 = 1$ for the common ancestor (see Fig.~\ref{fig:lineages}A). A lineage can become extinct in generation $n$ if $x_{n-1}$ has no offspring, ie.~$m_n = 0$.

We can relate properties of the population $\Psi$ to the statistical behaviour of individual lineages by means of the \emph{forward distribution} over lineages \cite{nozoe_inferring_2017}. The forward distribution is defined as follows: starting with the universal ancestor $x_0$, go down the population tree by picking a descendant uniformly at random in each generation. Since there are $m_i$ descendants to choose from in the $i$-th generation, each descendant has probability $m_i^{-1}$ of being picked. As a consequence, the forward probability of a lineage $\ell$ alive in generation $n$ equals
\begin{align}
    p_f(\ell = (x_0, x_1, \ldots, x_n) \given \Psi) = m_1^{-1} \cdots m_n^{-1}. \label{eq:pop_fwd_prob}
\end{align}

\noindent We stop this process when we reach an individual which has no offspring and the lineage becomes extinct. Here and in what follows, the subscript $f$ indicates the forward distribution, to be contrasted with the backward distribution introduced later.

From the forward distribution we recover the population size $N_n$ in the $n$-th generation as \cite{nozoe_inferring_2017}
\begin{align}
    N_n(\Psi) &= \E_f\left[ \prod_{i=1}^n m_i \, \big | \, \Psi \right]. \label{eq:pop_nozoe_N}
\end{align}

\noindent To show this, it is enough to consider all lineages up to the $n$-th generation. If a lineage goes extinct in generation $i \leq n$, then $m_i = 0$ and its total contribution to \eqref{eq:pop_nozoe_N} vanishes. Otherwise, its weight in Eq.~\eqref{eq:pop_nozoe_N} exactly cancels out its forward probability \eqref{eq:pop_fwd_prob}, and it contributes $1$ to the expectation.

Each lineage in the forward distribution evolves independently of the others in the population $\Psi$. Averaging over all possible realizations of a population $\Psi$, we can therefore define the forward lineage process, which describes the behaviour of a randomly sampled lineage irrespective of the surrounding population. Using Eq.~\eqref{eq:pop_nozoe_N}, we can write the \emph{expected} size of a population in generation $n$ as
\begin{align}
    \E[N_n] &= \E_f\left[ \prod_{i=1}^n m_i \right]. \label{eq:pop_nozoe_N_avg}
\end{align}

\noindent Here we sample a lineage from the forward lineage process, which is independent of $\Psi$ \cite{levien_large_2020}. Based on Eq.~\eqref{eq:pop_nozoe_N_avg} we define the weight, or absolute fitness, of a lineage $\ell$ as
\begin{align}
    w_n(\ell) &= \prod_{i=1}^n m_i. \label{eq:pop_abs_fitness}
\end{align}

\noindent This provides an intrinsic notion of reproductive success for a lineage \cite{nozoe_inferring_2017}. If a lineage goes extinct in generation $n$, then it does not contribute to future generations and we have $w_k(\ell) = 0$ for $k > n$.

This describes lineages in the $n$-picture, where the generations $n$ are fixed and the birth times $t_n$ are dependent variables. In the $t$-picture, we swap the roles of $n$ and $t$: lineages are now indexed by physical time $t$, and the current generation $n(t)$ is a dependent variable. We define the generation counting process $n(t)$ of a lineage as
\begin{align}
    n(t) &= i \qquad (t_i \leq t < t_{i+1}), \label{eq:pop_counting_process}
\end{align}

\noindent together with its weight process 
\begin{align}
    w(t) = w_{n(t)}.
\end{align}

\noindent This is illustrated in Fig.~\ref{fig:intro}B. The two representations are equivalent, and the $n$-picture and the $t$-picture of a lineage contain the same information.

The analogue of Eq.~\eqref{eq:pop_nozoe_N} at a fixed time $t$ is
\begin{align}
    N(t \given \Psi) &= \E_f\left[w(t) \given \Psi \right], \label{eq:pop_nozoe_T_tent}
\end{align}

\noindent where we still use the forward distribution defined by Eq.~\eqref{eq:pop_fwd_prob}. Eq.~\eqref{eq:pop_nozoe_T_tent} is only strictly valid for populations where individuals produce all their offspring at once and die (so-called splitting processes). This is the case e.g.~for cell division, but excludes models where individuals can procreate more than once. As we show in \ref{apdx:pop_count} however, Eq.~\eqref{eq:pop_nozoe_T_tent} is still \emph{asymptotically} valid for large $t$, which is enough for our purposes. We therefore arrive at the dual identities
\begin{align}
    \E[N_n] &= \E_f[w_n], & \E[N(t)] &\sim \E_f[w(t)], \label{eq:pop_asymp}
\end{align}

\noindent which relate population growth to the behavior of individual lineages. Here and in the rest of this paper, we use $\sim$ to signify that two quantities asymptotically grow at the same exponential rate: mathematically speaking, we have 
\begin{align}
    \quad N(t) &\sim e^{\Lambda t} &\LRA \lim_{t \rightarrow \infty} \frac 1 t \log \E[N(t)] &= \Lambda. \label{eq:sim_def}
\end{align}

\subsection*{The thermodynamic limit}

\label{sec:kappa}

A population consists of many interrelated lineages that evolve stochastically over time. If the population is large, we can treat it as a macroscopic object made up of microscopic particles (individual lineages), much like a gas that is made up of atoms or molecules. Since we are interested in the long-time behaviour of growing populations, this suggests that we can study a population from a thermodynamic perspective. More precisely, we can define two different ensembles of lineages: the $n$-ensemble, which consists of all lineages in a fixed generation $n$, and the $t$-ensemble, which consists of all lineages at a fixed time $t$. When generations overlap, there is no one-to-one relationship between $n$ and $t$, and the two ensembles will contain different lineages. Nevertheless, as we increase $n$ and $t$, both ensembles will eventually encompass all lineages in the population, and so we expect them to become equivalent in the long run. 

To make this statement more precise, we have to define what we mean by equivalence. In thermodynamics, an ensemble is typically defined by its partition function, which relates the statistical behaviour of the microscopic elements to macroscopic properties of the system. How can we define partition functions for the $n$- and $t$-ensembles? The most fundamental observable for both ensembles is their respective growth rate, Eq.~\eqref{eq:first}. In the $n$-ensemble, the only other natural variables are the generation times $t_n$; in the $t$-ensemble, it they are the generation counts $n(t)$. Based on this, we propose the definition 
\begin{align}
    \kappa_N(\alpha) &= \lim_{n\rightarrow \infty} \frac 1 n \log \E[w_n \, e^{-\alpha t_n}], \label{eq:pop_kappa_N}\\
    \kappa_T(\xi) &= \lim_{t \rightarrow \infty} \frac 1 t \log \E[w(t) \, e^{-\xi n(t)}], \label{eq:pop_kappa_T}
\end{align}

\noindent for the log-partition functions $\kappa_N$ and $\kappa_T$ of the two ensembles. Equivalently, the two functions are defined by the identities
\begin{align}
    \E[w_n \, e^{-\alpha t_n}] &\sim e^{\kappa_N(\alpha) \, n}, \label{eq:pop_kappa_N_impl}\\
    \E[w(t) \, e^{-\xi n(t)}] &\sim e^{\kappa_T(\xi) \, t}. \label{eq:pop_kappa_T_impl}
\end{align}

To motivate our definition, note that every lineage in the population is a possible fate of the original ancestor, and can be viewed as a ``state'' of our population. In this interpretation, $N(t)$ and $N_n$ count the number of states, and correspond to partition functions in thermodynamics. Eqs.~\eqref{eq:pop_kappa_N_impl} and \eqref{eq:pop_kappa_T_impl} measure how the number of states grows as a function of the parameters $\alpha$ and $\xi$, whence they can be seen as log-partition functions. For the moment, $\alpha$ and $\xi$ will be treated as a useful mathematical trick; we provide a physical interpretation at the end of this section. For $\alpha = 0$ and $\xi = 0$ we obtain
\begin{align}
    \kappa_N(0) &= \log R_0, & \kappa_T(0) &= \Lambda.
\end{align}

The function $\kappa_N$ becomes more familiar in the case where each individual has exactly $m$ offspring and generation lengths $\tau$ are independently sampled from a common distribution $f(\tau)$: 
\begin{align}
    \kappa_N(\alpha) &= \log \left( m \E[ e^{- \alpha \tau}] \right), \label{eq:kappa_N_indep}
\end{align}

\noindent which is the shifted logarithm of the Laplace transform of $f(\tau)$. We observe that the term in the logarithm directly appears in the classical Euler-Lotka equation \eqref{eq:el}. The Laplace transform is a mathematically equivalent way of representing the distribution $f(\tau)$: every property of the distribution can be recovered from $\kappa_N(\alpha)$. This principle, where the log-partition function encodes all relevant information about the ensemble, will hold more generally in what follows. 

Mathematically speaking, the log-partition functions $\kappa_N$ and $\kappa_T$ are rescaled Laplace transforms of the variables $t_n$ and $n(t)$, weighted by the multiplicity of each lineage to give population statistics following Eqs.~\eqref{eq:pop_nozoe_N_avg} and \eqref{eq:pop_nozoe_T_tent}. Much like ordinary Laplace transforms, $\kappa_N$ and $\kappa_T$ encode most of the relevant information about the underlying distributions, such as moments, in a convenient algebraic form. The scaling is chosen to give meaningful results in the limits $n \rightarrow \infty$ or $t \rightarrow \infty$ (compare Eq.~\eqref{eq:kappa_N_indep}), which define the asymptotic behaviour of the two ensembles. 

Since $\kappa_N$ and $\kappa_T$ encode the macroscopic behavior of our two ensembles, we can now formulate our thermodynamic equivalence in terms of these two functions. $\kappa_N$ and $\kappa_T$ are defined by the large-scale fluctuations of the dual variables $t_n$ and $n(t)$, which are directly related (Fig.~\ref{fig:intro}C): fluctuations in the birth times $t_n$ for fixed $n$ can equivalently be represented as fluctuations in the generation $n(t)$ for fixed $t$. A careful analysis of this relationship in \ref{apdx:ld}, based on the results in \cite{duffy_how_2005,pigolotti_generalized_2021}, yields the simple identity
\begin{align}
    \kappa_T(\kappa_N(\alpha)) &= \alpha. \label{eq:pop_duality}
\end{align}

\noindent In other words, $\kappa_N$ and $\kappa_T$ are inverse to each other. As these functions entirely determine the large-scale behaviour of the $n$-ensemble and the $t$-ensemble, Eq.~\eqref{eq:pop_duality} encapsulates the thermodynamic equivalence of our two ensembles. The rest of this paper is devoted to unravelling the consequences of this rather abstract identity.

Eq.~\eqref{eq:pop_duality} is equivalent to the reciprocity relation
\begin{align}
    \E[w_n \, e^{-\alpha t_n} ] \sim e^{\xi n} &\Leftrightarrow \E[w(t) \, e^{-\xi n(t))}] \sim e^{\alpha t}. \label{eq:pop_duality_reciproc}
\end{align}

\noindent A somewhat less rigorous mnemonic that illustrates the symmetric nature of this duality is
\begin{align}
    \E[w \cdot e^{-\alpha t - \xi n}] &\sim 1. \label{eq:pop_duality_mnem}
\end{align}

\noindent This illustrates a direct relationship between fluctuations in the quantity $n$ at a fixed time $t$, and the time it takes for $n$ to reach a certain value (a first passage time problem). The same formal relationship, ultimately based on the duality in \cite{glynn_large_1994,duffy_how_2005}, has recently been applied in a variety of thermodynamic contexts \cite{gingrich_fundamental_2017,raghu_thermodynamic_2025}.

To investigate the consequences of Eq.~\eqref{eq:pop_duality}, we start with the basic reproductive number $R_0$ and growth rate $\Lambda$, which are represented by
\begin{align}
    \log R_0 = \kappa_N(0) &\LRA \kappa_T(\log R_0) = 0, \label{eq:pop_R0_Lambda_quad_N} \\
    \Lambda = \kappa_T(0) &\LRA \kappa_N(\Lambda) = 0. \label{eq:pop_R0_Lambda_quad_T}
\end{align}

\noindent Written out, the last equation reads
\begin{align}
    \lim_{n\rightarrow \infty} \frac 1 n \log \E[w_n \, e^{-\Lambda t_n} ] &= 0,  \label{eq:pop_el_implicit}
\end{align}

\noindent which we recognize as a general form of the Euler-Lotka equation. Assuming that every organism has exactly $m$ descendants we obtain the version derived in \cite{pigolotti_generalized_2021},
\begin{align}
    \lim_{n\rightarrow \infty} \frac 1 n \log \E[e^{-\Lambda t_n}] &= -\log m,\label{eq:pop_el_implicit_const}
\end{align}

\noindent see \cite{cure_exponential_2025} for a further generalisation. The duality between $\kappa_N$ and $\kappa_T$ implies a direct relationship between the basic reproductive number $R_0$ and the growth rate $\Lambda$, and it explains why the Euler-Lotka equation is almost always encountered in implicit form: most population models are described in the $n$-picture, and the function $\kappa_N$ is more directly computable in practice.

Both $\kappa_N$ and $\kappa_T$ are decreasing convex functions where  $\log R_0$ and $\Lambda$ are their axis intercepts, see Fig.~\ref{fig:lineages}B. A physical interpretation of $\kappa_N$ (and similarly, $\kappa_T$) can be obtained as follows. Assume we modify our population model so that organisms are removed from the population at a constant rate $\eps_T > 0$; as shown in \ref{apdx:kappa_int}, this has the effect of shifting $\kappa_N$ to the left by an amount $\eps_T$ in Fig.~\ref{fig:lineages}B. The logarithm of the modified reproductive number is the new intercept with the vertical axis, which is equal to $\kappa_N(\eps_T)$. In other words, $\kappa_N$ measures how the average offspring per individual (the growth rate in the $n$-ensemble) changes when we modify the system in the $t$-picture. We can interpret the variable $\alpha$ in Eq.~\eqref{eq:pop_kappa_N} as a rate per unit time. Dually, $\kappa_T$ measures how the growth rate $\Lambda$ changes when we modify the system by an amount $\eps_N$ in the $n$-picture, as discussed in \ref{apdx:kappa_int}; here the parameter $\xi$ can be interpreted as a rate per generation.

The thermodynamic analogy can be made more concrete if we view the population growth rate $\Lambda$ as a form of internal pressure, being the propensity of the population to expand. Introducing a death rate $\eps_T$ as above can then be seen as applying an external force to the population. The population can maintain a steady-state precisely if the rate at which individuals are born equals the rate at which they are removed, ie.~$\Lambda = \eps_T$. In the $n$-picture, the population can maintain a steady state precisely if the average number of individuals per generation remains unchanged, ie.~$R_0(\eps_T) = 1$, or in other words, $\kappa_N(\Lambda) = 0$. This is the statement of the Euler-Lotka equation \eqref{eq:pop_el_implicit}. Again, a dual statement holds in the $n$-ensemble.

The thermodynamic equivalence of the $n$ and $t$-ensembles formally resembles the equivalence of the canonical and microcanonical ensembles in statistical physics. In the former, the system is taken at a fixed temperature, in the latter, it is taken at a fixed total energy. For a finite system, the canonical ensemble exhibits energy fluctuations around a mean value, so that its total energy cannot be clearly defined \textemdash{} the two ensembles are not equivalent. In the thermodynamic limit where the system size becomes infinite, the law of large numbers implies that these energy fluctuations become smaller and smaller, so that the system effectively behaves as if it had a fixed energy; asymptotically, it becomes equivalent to the microcanonical ensemble. In our paper, the situation is similar: if we fix the generation $n$, we observe fluctuations in $t_n$, and if we fix $t$, we observe fluctuations in $n(t)$. For long times, however, these fluctuations follow definite statistical patterns and we can directly translate from one ensemble to the other using Eq.~\eqref{eq:pop_duality}.

\subsection*{Individual histories}

\label{sec:bwd}

It is a fundamental observation in population biology that the history of a typical individual, which is encoded by its ancestral lineage or genealogy, is statistically distinct from a typical forward lineage: time is not reversible in a population \cite{georgii_supercritical_2003,jagers_stabilities_1992}. This phenomenon is well-known in population genetics \cite{nee_reconstructed_1994}, but applies very generally: lineages that reproduce faster, or have more offspring, will represent the majority of individuals, whereas lineages that have died out are absent. To understand the typical history of an individual in the population, we therefore have to understand the effect of selection acting within a population, which biases ancestral lineages towards fitter individuals.

The idea of sampling a random individual from a population $\Psi$ at a time $t$ can be formalised via the backward distribution over lineages \cite{georgii_supercritical_2003,nozoe_inferring_2017}. The backward distribution is given by
\begin{align}
    p_b(\ell \given \Psi) &= \frac 1 {N(t \given \Psi)} = \frac {w(t) \, p_f(\ell \given \Psi)} {N(t \given \Psi)}, \label{eq:pop_bwd_psi}
\end{align}

\noindent for any lineage alive at time $t$. The first equality assigns every individual at time $t$ equal weight; the second equality follows from the definition of the forward distribution, Eq.~\eqref{eq:pop_fwd_prob} and is illustrated in Fig.~\ref{fig:lineages}A. Here and in what follows, we will use the subscript $b$ to indicate the backward distribution as opposed to the forward distribution. 

Since the population $\Psi$ grows exponentially in time with fixed rate $\Lambda$, for large $t$ we can define the backward probability of any lineage without reference to the surrounding population as
\begin{align}
    p_b(\ell) &\propto w(t) \, e^{-\Lambda t} p_f(\ell). \label{eq:pop_bwd_T}
\end{align}

\noindent In the context of branching processes, the backward distribution is also known as the size-biased distribution or spinal decomposition \cite{joffe_exact_1982,lyons_conceptual_1995,georgii_supercritical_2003,olofsson_size-biased_2009} and forms a fundamental tool in the analysis of the long-term behavior of a process. The backward distribution weighs a lineage by its fitness $w(t)$ at a fixed time $t$, normalised by the asymptotic population size $e^{-\Lambda t}$. It represents the ancestral lineage of a typical lineage sampled from a population at large time $t$.

The description in the $t$-picture is natural e.g.~in biological experiments, where organisms are left to proliferate for a fixed duration and their historical fitness is measured using Eq.~\eqref{eq:pop_bwd_T} \cite{nozoe_inferring_2017}. If we want to model how changes in a population affect survival, it is more expedient to measure the fitness in each generation. We can accordingly define the backward distribution in the $n$-ensemble as
\begin{align}
    p_b(\ell) &\propto w_n \, e^{-\Lambda t_n} \, p_f(\ell). \label{eq:pop_bwd_N}
\end{align}

\noindent This determines how differences in generation lengths, measured by $t_n$, affect fitness (see Fig.~\ref{fig:lineages}C). Note that the growth rate $\Lambda$, which simply acts as a normalization constant in Eq.~\eqref{eq:pop_bwd_T}, determines the trade-off between time and fitness in the $n$-picture. Eqs.~\eqref{eq:pop_bwd_T} and \eqref{eq:pop_bwd_N} are not identical for finite $n$ or $t$ (they are defined for different sets of lineages), but they are asymptotically equivalent: both describe the history of a randomly sampled individual for sufficiently long times. 

The backward distribution is essential for understanding dynamical processes such as inheritance, mutation, and gene expression in populations. Modeling these along a typical forward lineage does not generally yield correct population statistics, since forward lineages are not representative of a population due to selection \cite{georgii_supercritical_2003,thomas_making_2017,hein_asymptotic_2024}. As we shall see in the next section, many quantities of interest such fitness gradients and selection coefficients are most naturally expressed in terms of backward lineages. Since  mathematical of populations are often formulated in the $n$-picture, Eq.~\eqref{eq:pop_bwd_T} allows us to compute these quantities in practice \textemdash{} in comparison, Eq.~\eqref{eq:pop_bwd_T} is often impractical as it ranges over individuals from many different generations.

As is commonly observed in thermodynamics, the log partition functions encode the statistics of ancestral lineages by their derivatives around the point $\kappa_T(0) = \Lambda$, see Fig.~\ref{fig:lineages}B. As shown in \ref{apdx:kappa_int}, differentiation of Eqs.~\eqref{eq:pop_kappa_N} and \eqref{eq:pop_kappa_T} yields
\begin{align}
    -\kappa_T'(0) &= \lim_{t\rightarrow \infty} \E_b\left[\frac{n(t)}{t} \right], \label{eq:pop_kappa_prime_bwd_T}\\
    -\kappa_N'(\Lambda) &= \lim_{n\rightarrow \infty} \E_b\left[\frac{t_n}{n} \right] := \E_b[\tau], \label{eq:pop_kappa_prime_bwd_N}
\end{align}

\noindent where the last term on the right denotes the mean generation length $\tau$ for backward lineages. The two identities are equivalent due to the law of large numbers,
\begin{align}
    \lim_{t\rightarrow \infty} \E_b\left[\frac{n(t)}{t} \right] &= \frac 1 {\E_b[\tau]}, \label{eq:pop_lln_bwd}
\end{align}

\noindent which directly follows from differentiating the fundamental identity Eq.~\eqref{eq:pop_duality}. The thermodynamic equivalence relation Eq.~\eqref{eq:pop_duality} can therefore be seen as a much more general statement of the law of large numbers, which predicts the average behaviour of lineages for large $n$ or $t$. Eq.~\eqref{eq:pop_duality} additionally yields e.g.~variances of generation lengths in a population via
\begin{align}
    \kappa_N''(\Lambda) &= \lim_{n\rightarrow \infty} \Var_b\!\left(\frac{t_n}{n} \right). \label{eq:pop_kappa_N_prime2_sigma}
\end{align}

\noindent This illustrates how the convexity of $\kappa_N$ and $\kappa_T$ is related to variability in the generation lengths of typical lineages \cite{yamauchi_unified_2022}. Indeed, if all generation lengths are equal, $\kappa_N$ and $\kappa_T$ become linear, and thermodynamic duality is trivial since $n$ and $t$ differ by a simple scaling factor (the length of a generation).

The convexity of $\kappa_N$ and $\kappa_T$, which is just Jensen's inequality in disguise, implies
\begin{align}
    \log R_0 &\geq \kappa_N(\Lambda) - \Lambda \kappa_N'(\Lambda) = \Lambda \E_b[\tau]. \label{eq:pop_bwd_el_ineq}
\end{align}

\noindent Thus the population as a whole grows slower in time than one would expect based on surviving lineages. This is not surprising if one considers that ancestral lineages are biased toward individuals that reproduce faster than the rest. A special case of Eq.~\eqref{eq:pop_bwd_el_ineq} for microbial models, together with an opposite-directed inequality for forward lineages, was derived in \cite{hashimoto_noise-driven_2016,garcia-garcia_linking_2019} in terms of the Kullback-Leibler divergence between the forward and backward distributions. Mathematically speaking, the backward distribution in Eq.~\eqref{eq:pop_bwd_N} can be seen an exponentially tilted version of the forward distribution, and the KL divergence can be computed directly in terms of the log-partition function $\kappa_N$. This provides a geometric way to understand this and similar inequalities \cite{genthon_fluctuation_2020,yamauchi_unified_2022} in terms of Bregman divergences. Note that the dual of Eq.~\eqref{eq:pop_bwd_el_ineq} for forward distributions does not hold for general population models as we show in \ref{apdx:counterexample}.

Our earlier description of $\kappa_N$ and $\kappa_T$ in terms of perturbations of the population can be used to obtain various sensitivity relations for the system. As shown in \ref{apdx:kappa_int}, removing a fraction $\eps_N \ll 1$ of newborn individuals from the population as they are born perturbs the growth rate as
\begin{align}
    \Lambda(\eps_N) &\approx \kappa_T(0) + \eps_N \kappa_T'(0) = \Lambda - \frac{\eps_N}{\E_b[\tau]} \label{eq:pop_Lambda_sens}
\end{align}

\noindent This expresses the change in growth rate in terms of the backward distribution; related sensitivity equations have been previously derived e.g.~in \cite{wakamoto_optimal_2012,yamauchi_unified_2022}. In an epidemic context, this predicts how the initial rate of spreading changes depending on the prevalence of immunity in the population. In the next section, we will extend Eq.~\eqref{eq:pop_Lambda_sens} to study how a population responds to more general perturbations, and see that the selection coefficient of a mutation can be computed in terms of backward lineages.


Our treatment of cumulant generating functions in this section is closely related to \cite{yamauchi_unified_2022}, and we derive more consequences of Eq.~\eqref{eq:pop_duality} in \ref{apdx:kappa_ext}, where we introduce an extension of the log partition functions $\kappa_N$ and $\kappa_T$ that encodes information about the asymptotic behavior of forward lineages, together with a generalisation of Eq.~\eqref{eq:pop_duality}. 

\subsection*{Selection coefficients}

We saw in the previous section that derivatives of the log-partition functions $\kappa_N$ and $\kappa_T$ encode population statistics such as mean generation lengths. We can use a similar approach to derive a general formula for the selection coefficient of a mutation, which quantifies fitness differences in population genetics \cite{chevin_measuring_2011}. This illustrates the central role of the backward distribution in understanding population dynamics.

Fix a wild-type population with growth rate $\Lambda$ and consider a perturbation which changes the growth rate as
\begin{align}
    \Lambda \mapsto \Lambda + \delta \Lambda. \label{eq:perturb_lambda}
\end{align}

\noindent As we show in \ref{apdx:sensitivity} following \cite{ilker_modeling_2019}, the selection coefficient for this mutation is approximately
\begin{align}
    s &\approx (\delta \Lambda) \E_b[\tau], \label{eq:selection_ilker}
\end{align}

\noindent where the second factor is the mean generation length of the original population, which is captured by the backward distribution. Eq.~\eqref{eq:selection_ilker} can be written as
\begin{align}
    s &\approx (\delta \kappa_N)(\Lambda),\label{eq:selection_kappa}
\end{align}

\noindent which is the first-order change in $\kappa_N$ for the mutant population, keeping $\Lambda$ fixed. We note that the selection coefficient $s$ is originally defined in the $n$-picture as the relative change in offspring per generation, while Eq.~\eqref{eq:selection_ilker} connects it to the $t$-picture via the physical growth rate $\Lambda$.

If the effect of the mutation can be written as a perturbation in offspring numbers and generation lengths of the form
\begin{align}
    \tau_i &\mapsto \tau_i + \delta \tau_i, \label{eq:perturb_tau} \\
    m_i &\mapsto m_i + \delta m_i,  \label{eq:perturb_m} 
\end{align}

\noindent then differentiating the generalised Euler-Lotka equation, Eq.~\eqref{eq:pop_el_implicit}, yields
\begin{align}
    s &\approx \E_b\left[ \frac{\delta m}{m} \right] - \Lambda \E_b[ \delta \tau ],\label{eq:selection_coefficient}
\end{align}

\noindent as we show in \ref{apdx:sensitivity}. Here the first term represents the mean relative change in offspring numbers per individual, and the second term is the mean change in generation length. Both averages are taken with respect to the backward distribution of the wild-type population. Eq.~\eqref{eq:selection_coefficient} reduces to well-known formul\ae{} e.g.~in \cite{charlesworth_evolution_1994,chevin_measuring_2011,ilker_modeling_2019} when there is no variability in generation times. We note that the averages in Eq.~\eqref{eq:selection_coefficient} can also be taken with respect to the mutant backward distribution, since switching the roles of the wild-type and mutant populations only changes the sign of all three terms in Eq.~\eqref{eq:selection_coefficient}.

Eq.~\eqref{eq:selection_coefficient} exhibits a fitness trade-off between offspring numbers and generation lengths, with the growth rate $\Lambda$ acting as a conversion factor. A mutation that reproduces faster, but has fewer offspring, will be selected for or against depending on the sign of Eq.~\eqref{eq:selection_coefficient}; the same applies to a mutation that produces more offspring, but later. 

Eq.~\eqref{eq:selection_coefficient} measures how a mutation affects the reproductive fitness of a typical lineage in the wild-type population. In other words, it approximates the behaviour of the perturbed population by means of the original population. This will break down if the mutation dramatically changes the population make-up, so that the backward lineages that contribute to the wild-type population become irrelevant in the mutant population. In this case, observable statistics of the backward distribution such as the mean generation length may change suddenly. Since such statistics are encoded by derivatives of the partition functions $\kappa_N$ and $\kappa_T$ (see e.g.~Eq.~\eqref{eq:pop_kappa_prime_bwd_N}), this corresponds to a scenario in which the partition functions change discontinuously \textemdash{} a thermodynamic phase transition.

\subsection*{Connection to branching processes}

\label{sec:appl}

We apply the theory developed above to general branching processes \cite{jagers_general_1989,athreya_branching_1972}, showcasing the concepts developed above with a view towards a simple epidemic model in the next section. Assume that each individual is assigned a type $x$ that encodes all relevant information about the organism \textemdash{} this could be its lifetime, its microbial phenotype, or its infectiousness in an epidemic. An individual of type $x$ produces a random number $m \geq 0$ of descendants, sampled from a probability distribution $p(m \given x)$. Conditioned on $m$, the probability that a descendant is of type $y$ and produced at age $\tau$ is given by the transition kernel $K_m(y, \tau \given x)$, unless of course $m = 0$, in which case there are no descendants.

The mean number of offspring of type $y$ produced by an individual of type $x$ throughout its lifetime defines the next-generation matrix \cite{diekmann_definition_1990}
\begin{align}
    M(y, x) &= \sum_{m \geq 1} p(m \given x) \, m \int_0^\infty K_m(y, \tau \given x) \, \dif \tau. \label{eq:appl_ngm}
\end{align}

\noindent We find (see \ref{apdx:branch})  that the mean number of offspring of type $y$ produced after $n$ generations is given by the $n$-th power, $M^n(y,x)$. Therefore, starting with an organism of type $x$, the expected population size after $n$ generations is given by
\begin{align}
    \E[N_n \given x] &= \sum_y M^n(y, x).
\end{align}

\noindent For large $n$, the behavior of $M^n$ is dictated by its dominant eigenvalue, which here coincides with its spectral radius $\spr(M)$, see \ref{apdx:perron}. We can therefore write asymptotically
\begin{align}
    \E[N_n \given x] &\sim \spr(M)^n. \label{eq:appl_ampl}
\end{align}

\noindent This shows that $R_0 = \spr(M)$ \cite{diekmann_definition_1990}. To obtain the log partition function $\kappa_N$ we introduce a constant death rate $\alpha > 0$ into our model, so that the average number of offspring of type $y$ produced by an individual of type $x$ becomes
\begin{align}
    M_{(-\alpha)}(y, x) &= \sum_{m \geq 1} p(m \given x) \, m \nonumber \\
    &\quad \qquad \cdot \int_0^\infty K_m(y, \tau \given x) \, e^{-\alpha \tau} \dif \tau. \label{eq:appl_M_tilt}
\end{align}

\noindent This differs from Eq.~\eqref{eq:appl_ngm} by an exponential term in the integral, which represents the probability that the parent survives to age $\tau$ (\ref{apdx:branch}). Our thermodynamic characterization of $\kappa_N$ together with the spectral radius formula imply
\begin{align}
    \kappa_N(\alpha) &= \log \spr(M_{(-\alpha)}). \label{eq:appl_kappa_N}
\end{align}

\noindent As a consequence, we obtain the matrix Euler-Lotka equation \cite{jagers_general_1989}
\begin{align}
    \spr(M_{(-\Lambda)}) &= 1. \label{eq:appl_el}
\end{align}

\noindent  In \ref{apdx:kappa_T} we compute $\kappa_T$ directly for this branching process and verify that Eq.~\eqref{eq:pop_duality} holds. For constant offspring numbers, Eq.~\eqref{eq:appl_el} can be derived directly from the results in \cite{pigolotti_generalized_2021} using the fact that the cumulant generating function for a Markov chain can be expressed in terms of the log spectral radius \cite{dembo_large_2010}.

The backward distribution over lineages, defined by Eqs.~\eqref{eq:pop_bwd_T} and \eqref{eq:pop_bwd_N}, is derived in \ref{apdx:branch_bwd}, where we show explicitly that the two definitions agree asymptotically. Backward lineages follow the Markov chain with transition kernel
\begin{align}
    K_b(y \given x) &= a_T(y) M_{(-\Lambda)}(y, x) \, a_T(x)^{-1}, \label{eq:appl_Kb}
\end{align}

\noindent where $a_T(x)$ is the reproductive value of an organism of type $x$, which coincides with the dominant left eigenvector of $M_{(-\Lambda)}$ (\ref{apdx:branch}). $K_b$ is a stochastic matrix since the corresponding dominant eigenvalue of $M_{(-\Lambda)}$ is $1$ by the Euler-Lotka equation \eqref{eq:appl_el}. 

For multitype branching processes we can define the population distribution $\pi_p(x)$ over types, called the tree population in \cite{nozoe_inferring_2017} and defined as the asymptotic distribution over all organisms produced in the population. As shown in \ref{apdx:branch}, the population distribution satisfies the equation
\begin{align}
    \pi_p(y) &= \sum_x M_{(-\Lambda)}(y, x) \, \pi_p(x), \label{eq:appl_dist_tree}
\end{align}

\noindent that is, $\pi_p$ is the right eigenvector of $M_{(-\Lambda)}$ with eigenvalue $1$. Perron-Frobenius theory (see \ref{apdx:perron}) and the matrix Euler-Lotka equation \eqref{eq:appl_el} guarantee that such an eigenvector exists. A version of the Euler-Lotka equation \cite{powell_growth_1956,lin_effects_2017,levien_interplay_2020} expresses the growth rate in terms of the population distribution of generation lengths and multiplicities
\begin{align}
    f_p(\tau, m) &= \sum_{x,y} p_m(m \given x) \, K_m(y, \tau \given x) \, \pi_p(x). \label{eq:appl_tau_marg_tree}
\end{align}

\noindent The growth rate $\Lambda$ then satisfies 
\begin{align}
    \int_0^\infty \sum_{m \geq 1} m \, f_p(\tau, m) \, e^{-\Lambda \tau} \dif \tau = 1. \label{eq:appl_el_tree}
\end{align}

\noindent This holds by Eq.~\eqref{eq:appl_dist_tree} if we note that $\pi_p$ is normalized to $1$ by assumption. Unfortunately, this is not a practical way to compute $\Lambda$: Eq.~\eqref{eq:appl_tau_marg_tree} requires the population distribution $\pi_p$, which in turn is given in terms of the growth rate $\Lambda$. Eq.~\eqref{eq:appl_el_tree} is analogous to formul\ae{} expressing $R_0$ in terms of the population distribution in epidemiology \cite{goltsev_localization_2012,cure_exponential_2025} in the $n$-picture.

\section*{Applications}

\subsection*{A simple epidemic model}

\label{sec:epi}

We illustrate the theory developed above with a simple epidemic model in the initial stages of an outbreak, using methods previously developed for microbial growth \cite{lin_effects_2017,lin_single-cell_2020,genthon_noisy_2025}. Assuming a large enough susceptible population, the initial stages of an epidemic can be represented as a branching process as discussed in the previous section. In the simplest case, every individual has a latency period $\tau$ following a distribution $f(\tau)$, after which it infects an average of $R_0$ other individuals, so that the spreading rate $\Lambda$ of the epidemic is given by the Euler-Lotka equation \eqref{eq:el}. This version of the equation treats every individual as equally infectious. We remove this assumption by introducing a strain-specific infectivity parameter $\iota$, such that the average offspring number is given by $e^\iota$, inherited according to the simple autoregressive model
\begin{align}
    \iota'  &= (1 - c_\iota) \overline{\iota} + c_\iota \iota + \sqrt{1 - c_\iota^2} \sigma_\iota \xi, \label{eq:epi_iotap_f}
\end{align}

\noindent with $\xi$ a unit normal random variable. Here \mbox{$0 \leq c_\iota < 1$} quantifies the heritability of infectiousness. The distribution of infectivities along a forward lineage has mean $\overline{\iota}$ and variance $\sigma_\iota^2$, regardless of heritability. 

The forward model defined in Eq.~\eqref{eq:epi_iotap_f} does not give an accurate account of epidemic spread as more infectious individuals contribute disproportionately to the population. We can verify that the population distribution over infectivities has mean
\begin{align}
    \E_p[\iota] &= \overline{\iota} + \frac{c_\iota}{1 - c_\iota} \sigma_\iota^2 \geq \E_f[\iota]. \label{eq:epi_iota_mean_p}
\end{align}

\noindent (Derivations for this and all following equations can be found in \ref{apdx:epi}.) The population is dominated by the offspring of successful individuals, which will be more infectious themselves as long as infectivities are heritable ($c_\iota > 0$). As a consequence, the net reproductive number is higher than expected based on forward lineages:
\begin{align}
    R_0 &= \E_p[e^\iota] = e^{\overline{\iota} + \frac{\sigma_\iota^2}{2} + \frac{c_\iota}{1 - c_\iota} \sigma_\iota^2} \geq \E_f[e^\iota].\label{eq:epi_r0}
\end{align}

\noindent Heritability increases the spread of the epidemic as more successful strains pass their advantage on to descendants.

\begin{figure*}[t]
    \centering
    \includegraphics{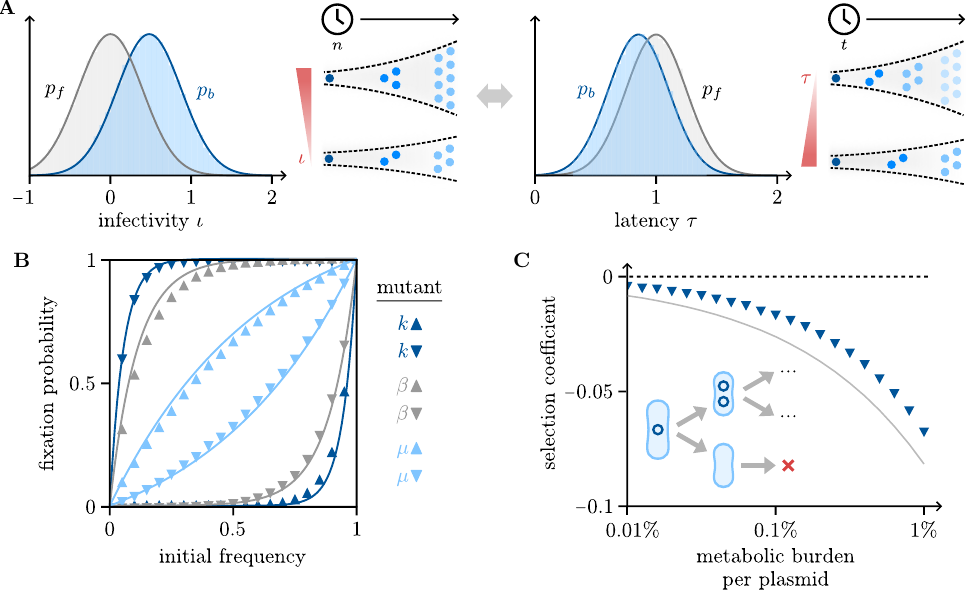}
    \caption{\textbf{A} Spread of an epidemic in the $n$-picture and the $t$-picture. On the left, strains with higher infectivity create more offspring per generation, and are overrepresented in backward lineages as measured by the fitness relation \eqref{eq:pop_bwd_T}. Theoretical predictions (solid lines) match simulations (histograms) for the forward distribution $p_f$ and the backward distribution $p_b$ over infectivities computed using Eq.~\eqref{eq:pop_bwd_N}. On the right, strains with lower latency propagate faster and are similarly overrepresented in backward lineages. The corresponding fitness relation is given by Eq.~\eqref{eq:pop_bwd_N}. \textbf{B} Fixation probabilities in a simple haploid population model. We simulated a wild-type red squirrel population with fixed carrying capacity $N = 100$ together with six different perturbations and estimated the fixation probability of the invading species numerically (triangles) and using Eqs.~\eqref{eq:selection_ilker} and \eqref{eq:p_fix} (lines). \textbf{C} Fitness cost of an artificial plasmid as a function of the per-plasmid metabolic burden $\beta$. We simulated the model  for various values of $\beta$ and compared their selection coefficient using the definition in Eq.~\eqref{eq:selection_ilker} (lines) and with the approximation in Eq.~\eqref{eq:selection_coefficient} (triangles). Simulation details for all three models can be found in \ref{apdx:models}. }

    \label{fig:appl}
\end{figure*}

In the long run, most infections arise from a few highly infectious individuals. If we were to trace the ancestry of infected individuals, we would observe a statistical pattern for infectivities that differs from Eq.~\eqref{eq:epi_iotap_f}. Indeed, the ancestry of an individual would appear to follow the autoregressive process
\begin{align}
    \iota' &= (1 - c_\iota) \overline{\iota} + (1 + c_\iota) \sigma_\iota^2 + c_\iota \iota + \sqrt{1 - c_\iota^2} \sigma_\iota \xi, \label{eq:epi_iotap_a}
\end{align}

\noindent which differs from Eq.~\eqref{eq:epi_iotap_f} by the second term. Thus ancestral infectivities are systematically biased upwards compared to forward lineages (see Fig.~\ref{fig:appl}A). This is a general effect of selection: highly infectious individuals have more offspring, even in the absence of heritability.

So far we have seen how differences in offspring numbers affect the fitness of lineages, working entirely in the $n$-picture. Following Eq.~\eqref{eq:pop_bwd_N} we can use an analogous approach to evaluate the effect of differences in transmission speeds. Assume for example that different strains have different latency periods $\tau$, inherited according to the autoregressive model
\begin{align}
    \tau' &= (1 - c_\tau) \overline{\tau} + c_\tau \tau + \sqrt{1 - c_\tau^2} \sigma_\tau \xi, \label{eq:epi_lambda_f}
\end{align}

\noindent independent of the infectivity. Here we assume small $\sigma_\tau$, so that $\tau' > 0$ in most cases. The growth rate becomes
\begin{align}
    \Lambda
    &= \frac {2 \log R_0} {\overline{\tau} + \sqrt{\overline{\tau}^2 - 2 \frac{1 + c_\tau}{1 - c_\tau} \sigma_\tau^2 \log R_0}}, \label{eq:epi_el} 
\end{align}

\noindent cf.~\cite{lin_single-cell_2020}. By Eq.~\eqref{eq:pop_bwd_N}, asymptotically we can treat differences in generation lengths exactly like differences in offspring numbers, with the growth rate $\Lambda$ acting as the conversion factor. Indeed, a calculation shows that the population average over latencies is given by
\begin{align}
    \E_p[\tau] &= \overline{\tau} - \Lambda \frac{c_\tau}{1 - c_\tau} \sigma_\tau^2 \leq \E_f[\tau].
\end{align}

\noindent This is entirely analogous to Eq.~\eqref{eq:epi_iota_mean_p}. If the latency period is heritable, ie.~$c_\tau > 0$, the population is biased towards strains with a shorter latency period. As with infectivities, backward lineages predominantly feature individuals with shorter latency periods. The evolution of latency periods in backward lineages is given by
\begin{align}
    \tau' &= (1 - c_\tau) \overline{\tau} - \Lambda (1 + c_\tau) \sigma_\tau^2 + c_\tau \tau + \sqrt{1 - c_\tau^2} \sigma_\tau \xi. \label{eq:epi_lambda_a}
\end{align}

\noindent We showcase the duality in this example in Fig.~\ref{fig:appl}A, where we numerically verify the predictions in this section. 

\subsection*{Invasion analysis}



We can use Eq.~\eqref{eq:selection_ilker} for the selection coefficient to predict the likelihood that a mutation will eventually reach fixation. Assuming a constant carrying capacity $N$, the fixation probability of a mutation can be computed using Kimura's formula\cite{kimura_probability_1962,alexandre_bridging_2025}:
\begin{align}
    p_{\mathrm{fix}} &= \frac{1 - e^{-N s f}}{1 - e^{-N s}}, \label{eq:p_fix}
\end{align}

\noindent where $f$ is the fraction of mutants in the original population. Here we consider an asexually reproducing haploid population, modeled as a Moran process to account for overlapping generations (in contrast to a Fischer-Wright process, which assumes fixed generation lengths). While Eq.~\eqref{eq:p_fix} assumes a fixed population size $N$ across generations, in contrast to our original framework, we will see that this does not affect the accuracy of our predictions.

We validate Eq.~\eqref{eq:p_fix} using a simple model of squirrel populations in the UK. Here, red squirrels (wild-type) compete against gray squirrels (introduced). We assume that the two populations do not interbreed and only consider female individuals, resulting in an effectively haploid population. We fix a carrying capacity $N$ and simulate individuals reproducing independently; whenever the population size exceeds $N$ following a reproduction event, we sample $N$ individuals at random from the population and continue. In our model, red squirrels reproduce on average once every six months with a standard deviation of one month, represented as a Gamma distribution. Each squirrel then gives birth to a geometrically distributed number of offspring with mean $2$. For simplicity, we assume that generation times and offspring sizes are independent across generations and neglect age effects. Since squirrels can reproduce over multiple rounds, we also treat the parent as an additional offspring; a mathematical description of our model can be found in \ref{apdx:sensitivity}. We then define six hypothetical mutants that differ in their offspring and generation time distributions and compare the predictions of Eqs.~\eqref{eq:selection_ilker} and \eqref{eq:p_fix} with numerical simulations in Fig.~\ref{fig:appl}B.

\subsection*{Plasmid engineering}

Bacterial growth rates are of fundamental importance in precision fermentation, which uses engineered microbes to produce e.g.~recombinant proteins at industrial scale. This is normally achieved by expressing heterologous proteins in bacteria using plasmids which are replicated and passed on across generations \cite{rugbjerg_synthetic_2018,ow_global_2006}. A fundamental problem here is that these plasmids interfere with the normal bacterial life cycle, not least due the metabolic burden incurred by the expression of target proteins \cite{glick_metabolic_1995,ilker_modeling_2019}, which leads to slowed growth of plasmid-bearing cells and thus a selective disadvantage. 

Consider a simple model of \textit{E.~coli} bearing several copies of a heterologous plasmid that are duplicated and inherited by its descendants. A cell inheriting $k$ plasmids from its parent will pass on $2k$ plasmids to its offspring, which segregate approximately at random \cite{hernandez-beltran_segregational_2022}. If the phenotype of a cell is determined by the number $k$ of plasmids inherited at birth, we can model this with a binomial transition kernel
\begin{align}
    K(k' \given k) &= \binom{k'}{2k} \, 2^{-2 k},  \label{eq:plasmid_K}
\end{align}

\noindent where $k$ and $k'$ are the number of plasmids inherited by the parent and daughter cell, respectively. We assume that each plasmid incurs a metabolic cost that modulates the time to division as
\begin{align}
    \tau(k) &= \tau_0 (1 + \beta k),\label{eq:plasmid_tau}
\end{align}

\noindent where $\beta$ measures the metabolic burden per plasmid. We neglect other sources of variation in interdivision times.

Eq.~\eqref{eq:plasmid_K} implies that the state $k = 0$ is absorbing: cells cannot regain plasmids once lost in the absence of horizontal gene transfer \cite{smillie_mobility_2010}. Since cells containing no plasmids are selectively favoured by Eq.~\eqref{eq:plasmid_tau}, this will eventually drive plasmid-containing cells to extinction. To prevent this unfavourable scenario, we implement the addiction mechanism in \cite{rugbjerg_synthetic_2018} where essential host genes are moved to the plasmid, which prevents cells with $k = 0$ from reproducing entirely. 

Predicting the growth rate analytically for this model is challenging, since the potentially unbounded number of plasmids per cell results in an infinite-dimensional Euler-Lotka equation \eqref{eq:el}. An increasing metabolic burden $\beta$ not only decreases the growth rate of cells via Eq.~\eqref{eq:plasmid_tau}, but also shifts the distribution of plasmid numbers in the population to lower values due to the increased fitness penalty for large $k$. However, we can use this model to verify Eq.~\eqref{eq:selection_coefficient}, which predicts the total fitness penalty incurred by the plasmids together with the addiction mechanism from \cite{rugbjerg_synthetic_2018}. 

To do so, we have to consider how either perturbation affects the terms in Eq.~\eqref{eq:selection_coefficient}. For a cell with $k$ plasmids, the expected relative change in offspring equals roughly
\begin{align}
    \E\left[\frac{\delta m}{m} \, \big| \, k\right] &\approx -2^{-2k}, \label{eq:plasmid_delta_m}
\end{align}

\noindent since the addiction mechanism effectively removes descendants with $k' = 0$ plasmids, the expected fraction of which equals $2^{-2k}$. The second term in Eq.~\eqref{eq:selection_coefficient} on the other hand yields
\begin{align}
    \Lambda (\delta \tau) &= -(\log 2) \beta k, \label{eq:plasmid_delta_tau}
\end{align} 

\noindent for a cell with $k$ plasmids, since $\Lambda = (\log 2)/\tau_0$ for the unperturbed model, cf.~\cite{ilker_modeling_2019}. We then average Eq.~\eqref{eq:plasmid_delta_m} and Eq.~\eqref{eq:plasmid_delta_tau} over the backward distribution for the \emph{perturbed} population with $\beta > 0$ \textemdash{} this is because the unperturbed model predicts arbitrarily large plasmid abundances in a population, which is not biological (\ref{apdx:plasmids}). Doing this we obtain an approximation for the selection coefficient, visualized in Fig.~\ref{fig:appl} for realistic values of $\beta \approx 0.01 - 1\%$ \cite{scott_interdependence_2010,rouches_plasmid_2022}. We remark that the approximation can be improved by taking into account second-order corrections to Eq.~\eqref{eq:plasmid_delta_m}, which we omit in this paper. 

\section*{Discussion}


We introduce a general thermodynamic framework for populations and lineages that relates the generational description of branching processes with their description in physical time via Eq.~\eqref{eq:pop_duality}. Our approach is inspired by recent work analyzing populations in terms of individual lineages~\cite{wakamoto_optimal_2012,sughiyama_pathwise_2015,levien_large_2020, yamauchi_unified_2022}, and is directly based on the duality between the processes $t_n$ and $n(t)$ uncovered in \cite{glynn_large_1994,duffy_how_2005,gingrich_fundamental_2017,pigolotti_generalized_2021}. This duality provides a new and unifying perspective on many results in the study of population growth, in particular the Euler-Lotka equation as well as the ancestry history of populations \cite{georgii_supercritical_2003,wakamoto_optimal_2012}. The latter is of major importance for modeling processes such as gene expression and cell size homeostasis in microbial populations, which depend on the history of individual lineages \cite{thomas_making_2017,thomas_intrinsic_2019,genthon_analytical_2022,hein_asymptotic_2024,ocal_cell_2025}. We apply our formalism to derive new formul\ae{} for the selection coefficient that allow us to estimate fixation probabilities and fitness differences in population genetics. Our results hold for branching processes with variable numbers of offspring, including the possibility of extinction \cite{genthon_cell_2023}, which is a common feature in epidemic and microbial models. 

Our thermodynamical framework is intrinsically asymptotic and does not apply exactly for finite populations. Nevertheless, our numerical examples show that this approach can yield accurate predictions for well-mixed populations on the order of $100-1000$ individuals. Experimental measurements in \cite{wakamoto_optimal_2012,hashimoto_noise-driven_2016,yamauchi_unified_2022} suggest that asymptotic considerations can provide remarkably accurate explanations of microbial dynamics. In epidemiology, it is well-known that the branching process approximation of an epidemic is only valid in the early stages of an epidemic. Our approach conceptually clarifies some long-term properties of branching processes, but the relationship between the $n$-ensemble and the $t$-ensemble in the non-asymptotic regime remains to be studied.

While the results of this paper do not make strong assumptions on the underlying population process, they fundamentally rely on the large deviation principle for lineages, the validity of which can be difficult to establish for complex models. Our approach is only valid if fluctuations in lineages are not too large as formalised by the Kesten-Stigum theorem \cite{lyons_conceptual_1995}, which ensures that the behaviour of lineages is captured by the appropriate exponential averages; this e.g.~excludes cases with very heavy-tailed offspring numbers. Large deviations must also be treated with care in the presence of stochastically fluctuating environments, which is particularly important for modeling realistic populations, and establishing general principles as outlined in \cite{kobayashi_fluctuation_2015,hein_asymptotic_2024} for such population processes remains a problem of major interest. 

\section*{Resource availability}

\subsection*{Lead contact}

Requests for further information and resources should be directed to and will be fulfilled by the lead contact, Kaan \"Ocal (kaan.ocal@unimelb.edu.au).

\subsection*{Materials availability}

This study did not generate new unique reagents.
    
\subsection*{Data and code availability}

All original code is publicly available at \url{https://github.com/kaandocal/twoclock}. Any additional information required to reanalyze the data reported in this paper is available from the lead contact upon request.

\subsection*{Acknowledgments}
The authors would like to thank Yong See Foo, Augustinas Sukys and the anonymous referees for feedback on the manuscript, and gratefully acknowledge financial support through an ARC Laureate Fellowship to MPHS (FL220100005).

\subsection*{Declaration of Interests}

MPHS serves on the advisory board of Cell Systems. MPHS is co-founder, shareholder, director, and CSO of Cell Bauhaus Pty Ltd.


%

\printbibliography

\clearpage
\onecolumn
\numberwithin{equation}{section}

\appendix

\renewcommand \thesection {Appendix~\Alph{section}}

\section{Lineages and populations}

\FloatBarrier

\label{apdx:pop_count}

\setcounter{figure}{0}
\renewcommand*\theequation{\textup{\Alph{section}\arabic{equation}}}

\renewcommand{\thefigure}{A\arabic{figure}}

\begin{figure}[t]
    \begin{minipage}{0.5\linewidth}
        \centering
        \includegraphics{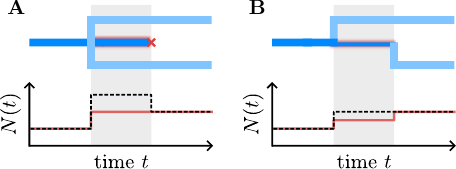}
    \end{minipage}~\begin{minipage}{0.48\linewidth}
    \caption{Failure modes for Eq.~\eqref{eq:pop_nozoe_T_tent}. \textbf{A} After producing all its offspring, the fate of an individual is not captured by the forward distribution, which underestimates the real population size. \textbf{B} If an individual procreates in multiple rounds, it will only be partially counted after the first offspring. The solid line indicates the prediction of Eq.~\eqref{eq:pop_nozoe_T_tent}, the dashed line the true population size. For growing populations, these differences are asymptotically negligible.}
    \label{fig:apdx_lineages}
    \end{minipage}
\end{figure}

Since the forward distribution over lineages is defined in terms of generations, there are two ways in which Eq.~\eqref{eq:pop_nozoe_T_tent} can fail for finite $t$. Consider the case where an organism $x_n$ has produced all its offspring and is still alive at time $t$ (see Fig.~\ref{fig:apdx_lineages}). By our definition of the forward distribution, all forward lineages involving $x_n$ will have moved to the $(n+1)$-st generation by time $t$, so the contribution of $x_n$ is not counted in \eqref{eq:pop_nozoe_T_tent}. This however does not affect the asymptotics for a growing population. Indeed, the population grows with asymptotic rate $\Lambda > 0$ precisely if the number of newborn individuals grows asymptotically as $e^{\Lambda t}$, so we can ignore organisms that have fulfilled their reproductive role for counting purposes.

Another problem appears if individuals can have offspring at different times. If the organism $x_n$ has offspring before and after time $t$, the contributions of the lineages that split off after time $t$ will not sum to $1$ under \eqref{eq:pop_nozoe_T_tent}, see Fig.~\ref{fig:apdx_lineages}. As a consequence, our formula may underestimate the contribution of $x_n$ by a factor of up to $m_{n+1}$. If we assume that the maximum number of offspring is bounded, Eq.~\eqref{eq:pop_nozoe_T_tent} underestimates the true population size by at most a constant factor, which again does not affect its asymptotic growth. More generally, as long as the offspring distribution decays sufficiently quickly, Eq.~\eqref{eq:pop_nozoe_T_tent} remains asymptotically valid. We note that the offspring distribution must decay sufficiently quickly for the Kesten-Stigum theorem to hold more generally \cite{lyons_conceptual_1995}.

\section{Large deviations}

\label{apdx:ld}

In this section we recapitulate the basics of large deviation theory required for the main argument of this paper, referring to \cite{dembo_large_2010} for a more complete and rigorous treatment. A sequence $X_1, X_2, \ldots$ of random variables satisfies a large deviation principle with rate function $I(x)$ if
\begin{align}
    p\left(\frac{X_n}{n} = x\right) &\sim e^{-n I(x)} \label{eq:apdx_ld_asymp}
\end{align}

\noindent for all $x$, which is equivalent to
\begin{align}
    \lim_{n \rightarrow \infty} \frac 1 n \log p\left(\frac{X_n}{n} = x\right) &= -I(x).
\end{align}

\noindent The limiting cumulant generating function of the sequence is defined as
\begin{align}
    \kappa(\lambda) &= \lim_{n \rightarrow \infty} \frac 1 n \log \E[e^{\lambda X_n}]. \label{eq:apdx_ld_cgf}
\end{align}

\noindent As the limit of convex functions, $\kappa$ is convex; if furthermore the $X_i$ are all positive, $\kappa$ is weakly increasing. Under suitable regularity conditions, Varadhan's Lemma states that $\kappa$ is the Legendre transform of the rate function $I$, that is,
\begin{align}
    \kappa(\lambda) &= \sup_x \lambda x - I(x). \label{eq:apdx_ld_cgf_legendre}
\end{align}

\noindent This formally follows by plugging Eq.~\eqref{eq:apdx_ld_asymp} into the expectation in Eq.~\eqref{eq:apdx_ld_cgf}:
\begin{align}
    \kappa(\lambda) &= \lim_{n \rightarrow \infty} \frac 1 n \log \int e^{n \lambda x} p\left(\frac{X_n}{n} = x\right) \dif x = \lim_{n \rightarrow \infty} \frac 1 n \log \int e^{n (\lambda x - I(x))} \dif x = \sup_x  \lambda x - I(x).
\end{align}

\noindent In the last step we used Laplace's method to approximate the integral in the limit of large $n$ \textemdash{} the integral will be dominated by the value that maximizes the exponent, as the relative contribution of other values will be suppressed exponentially in $n$. If the rate function is convex, Legendre duality implies that we can recover $I$ from $\kappa$ as
\begin{align}
    I(x) &= \sup_\lambda \lambda x - \kappa(\lambda). \label{eq:apdx_ld_rf_legendre}
\end{align}

\noindent The above principles still apply if the discrete parameter $n$ is replaced by a continuous parameter $t$.

As an example we can consider the case where 
\begin{align}
    X_n &= Z_1 + Z_2 + \ldots + Z_n,
\end{align}

\noindent where the $Z_k$ are independent identically distributed random variables. Then
\begin{align}
    \E[e^{\lambda X_n}] &= \E[e^{\lambda Z}]^n,
\end{align}

\noindent so $\kappa$ is just the cumulant generating function of the $Z_k$. This applies to a population model where intergeneration times are independent and identically distributed: $\kappa_N$ becomes the cumulant generating function of this distribution.

The definition of the counting process $n(t)$ associated to the point process $t_n$ in Eq.~\eqref{eq:pop_counting_process} implies a certain relationship between their rate functions \cite{glynn_large_1994,gingrich_fundamental_2017}. To extend the argument in \cite{glynn_large_1994,pigolotti_generalized_2021} to a lineage process with variable offspring numbers we follow the general argument presented in \cite{duffy_how_2005,gingrich_fundamental_2017}. Keeping track of the variables $h_n = \log w_n$ and $h(t) = \log w(t)$ we can compute compute
\begin{align}
    e^{-n I_N(\tau, \eta)} \approx p\left(\frac{t_n}{n} = \tau, \frac{h_n}{n} = \eta\right) \approx p\left(\frac {n(t)}{t} = 1/\tau, \frac{h(t)}{t} = \eta/\tau\right) \approx e^{-t I_T(1/\tau, \eta/\tau)}
\end{align}

\noindent for large $n$, where we substitute $t = n \tau$. This implies that
\begin{align}
    I_N(\tau, \eta) &= \tau \, I_T(1/\tau, \eta/\tau). \label{eq:apdx_ld_duality_rd}
\end{align}

\noindent Introduce the two-dimensional cumulant generating functions
\begin{align}
    \tilde \kappa_N(\alpha, \beta) &= \lim_{n \rightarrow \infty} \frac 1 n \log \E[ e^{\beta h_n - \alpha t_n}], & \tilde \kappa_T(\xi, \beta) &= \lim_{t \rightarrow \infty} \frac 1 t \log \E[ e^{\beta h(t) - \xi n(t)}]. \label{eq:apdx_ld_cgf2}
\end{align}


\noindent Upon taking Legendre transforms, Eq.~\eqref{eq:apdx_ld_duality_rd} implies
\begin{align}
    \tilde \kappa_N(\alpha, \beta) &= \sup_{\tau,\eta} \beta \eta - \alpha \tau - I_N(\tau, \eta) = \sup_{\tau,\eta} \tau \left(\beta \frac {\eta}{\tau} - \alpha - I_T\left(\frac 1 \tau, \frac {\eta} {\tau}\right)\right) = \sup_{\tau,\tilde \eta} \tau \left(\beta \tilde \eta - \alpha - I_T\left(\frac 1 \tau, \tilde \eta\right)\right) \nonumber \\
    &= \sup_{\tau,\tilde \eta} \; \inf_{\xi,\gamma} \tau \left(\beta \tilde \eta - \alpha - \gamma \tilde \eta + \frac {\xi} \tau - \tilde \kappa_T(\xi, \gamma)\right).
\end{align}

\noindent We next invoke the minimax principle to swap the order of supremum and infimum:
\begin{align}
    \tilde \kappa_N(\alpha, \beta) &= \inf_{\xi,\gamma} \; \sup_{\tau,\tilde \eta} \xi + \tau \left(\beta \tilde \eta - \alpha - \gamma \tilde \eta - \tilde \kappa_T(\xi, \gamma)\right).
\end{align}

\noindent If the term in brackets is nonzero, the supremum is infinite, which follows from choosing $\tau$ large and of the same sign. The infimum of the whole expression is therefore determined by the vanishing of the bracketed term, which implies
\begin{align}
    \gamma &= \beta, & \alpha &= \tilde \kappa_T(\xi, \beta),
\end{align}

\noindent and we obtain the duality relation
\begin{align}
    \tilde \kappa_N(\alpha, \beta) = \xi &\LRA \tilde \kappa_T(\xi, \beta) = \alpha, \label{eq:apdx_ld_duality_ext}
\end{align}

\noindent as shown in \cite{gingrich_fundamental_2017}. For fixed $\beta$, $\tilde \kappa_N$ and $\tilde \kappa_T$ are inverse to each other, which generalizes the result in \cite{pigolotti_generalized_2021}. Now Eq.~\eqref{eq:pop_duality} follows by taking $\beta = 1$. Our derivation is a simplified version of \cite{duffy_how_2005}, whose rate functions $I$ and $J$ correspond to $I_N$ and $I_T$, respectively. Note that our $n$-ensemble corresponds to the $\delta$-ensemble in \cite{pigolotti_generalized_2021} (the division ensemble). 

This duality argument can be explicitly verified when $n(t)$ follows a Poisson process with constant rate $\lambda$. For simplicity we set $\beta = 0$. Then $n(t) \sim \mathrm{Poi}(\lambda t)$ and we obtain
\begin{align}
    \kappa_T(\xi) &= \lambda (e^{-\xi} - 1).
\end{align}

\noindent The waiting times for this process are independent samples from $\mathrm{Exp}(\lambda)$, so $t_n$ follows an Erlang distribution and we can compute directly that
\begin{align}
    \kappa_N(\alpha) &= -\log \left(1 + \frac {\alpha}{\lambda}\right).
\end{align}

\noindent These two functions inverse to each other \cite{glynn_large_1994,gingrich_fundamental_2017}. 

\section{Interpreting the partition function}

\label{apdx:kappa_int}

Consider a modification of our population model where organisms die with constant rate $\eps_T > 0$. Then the probability that a lineage $\ell$ in the original model is still alive at time $t$ is $e^{-\eps_T t}$, so the modified partition function $\kappa_T^{(\eps_T)}$ is given by
\begin{align}
    \kappa^{(\eps_T)}_T(\xi) &= \lim_{t \rightarrow \infty} \frac 1 t \log \E[w(t) \, e^{-\xi n(t)-\eps_T t}] = \kappa_T(\xi) - \eps_T.
\end{align}

\noindent It follows from \eqref{eq:pop_duality} that
\begin{align}
    \kappa^{(\eps_T)}_N(\alpha) &= \kappa_N(\alpha + \eps_T),
\end{align}

\noindent which is equivalent to shifting the graph in Fig.~\ref{fig:lineages} to the left by $\eps_T$.

Now consider an alternative modification of our population model where only a fraction $e^{-\eps_N}$ of individuals born remains in the population (in an epidemic, this can be seen as the fraction of susceptible individuals). The probability that a lineage $\ell$ in the original model is still alive in generation $n$ is $e^{-n \eps_N}$, so the modified partition function $\kappa_N^{(\eps_N)}$ is given by
\begin{align}
    \kappa^{(\eps_N)}_N(\alpha) &= \lim_{n \rightarrow \infty} \frac 1 n \log \E[w_n \, e^{-\alpha t_n - n \eps_N}] = \kappa_N(\alpha) - \eps_N,
\end{align}

\noindent which is equivalent to shifting the graph in Fig.~\ref{fig:lineages} down by $\eps_N$. In particular, the basic reproductive number becomes $e^{-\eps_N} R_0 < R_0$ and the new growth rate is given by the solution of
\begin{align}
    \kappa^{(\eps_N)}_T(\Lambda(\eps_N)) &= \kappa_T(\Lambda(\eps_N) + \eps_N) = 0.
\end{align}

\noindent To first order in $\eps_N$, the growth rate changes as indicated in Eq.~\eqref{eq:pop_Lambda_sens}. 

We can differentiate Eqs.~\eqref{eq:pop_kappa_N} and \eqref{eq:pop_kappa_T} to obtain
\begin{align}
    -\kappa'_N(\Lambda) &= \lim_{n \rightarrow \infty} \frac{\E\left[w_n \, e^{-\Lambda t_n} \left(\frac{t_n}{n}\right)\right]}{\E[w_n \, e^{-\Lambda t_n}]}, \\
    \kappa''_N(\Lambda) &= \lim_{n \rightarrow \infty} \left( \frac{\E\left[w_n \, e^{-\Lambda t_n} \left(\frac{t_n^2}{n}\right)\right]}{\E[w_n \, e^{-\Lambda t_n}]} - \frac{\E\left[w_n \, e^{-\Lambda t_n} \left(\frac{t_n}{n}\right)\right]^2}{\E[w_n \, e^{-\Lambda t_n}]^2} \right), \\
    -\kappa'_T(0) &= \lim_{t \rightarrow \infty}\frac{\E\left[w(t) \frac{n(t)}{t} \right]}{\E\left[w(t)\right]}.
\end{align}

\noindent These can be interpreted as moments of intergeneration times for the backward distribution $p_b$ since
\begin{align}
    p_b(\ell) &= \frac {w_n \, e^{ - \Lambda t_n}} {\E[w_n \, e^{-\Lambda t_n}]} \, p_f(\ell)
\end{align}

\noindent in the $n$-ensemble and
\begin{align}
    p_b(\ell) &= \frac {w(t)} {\E[w(t)]} \, p_f(\ell) 
\end{align}

\noindent in the $t$-ensemble. 

\section{The extended partition function}

\label{apdx:kappa_ext}

The functions $\kappa_N$ and $\kappa_T$ encode information about the asymptotic behavior of a population, which is determined by the behavior typical ancestral lineages. In this section we will define extended log partition functions that allow us to analyze the asymptotic behavior of both forward and backward lineages together, and use this to deduce additional properties of the population model similar to \cite{yamauchi_unified_2022}.

To analyze the behavior of forward lineages, we need to treat the possibility of death more carefully. Eq.~\eqref{eq:pop_kappa_N} ignores the lifetime of individuals after procreation, and in particular, that of individuals with no offspring. In \ref{apdx:pop_count} we argue that this is irrelevant for the asymptotic behavior in a growing population, but in this section we will assume that organisms post reproduction have a finite maximal lifetime $T_d$ (more generally, it suffices that their residual lifetime distribution decays sufficiently quickly). 

We therefore define the extended log partition with additional parameter $\beta > 0$ as
\begin{align}
    \tilde \kappa_N(\alpha, \beta) &= \lim_{n\rightarrow \infty} \frac 1 n \log \E[w_n^\beta \, e^{-\alpha t_n} ], & \tilde \kappa_T(\xi, \beta) &= \lim_{t \rightarrow \infty} \frac 1 t \log \E[w(t)^\beta \, e^{-\xi n(t)} ].\label{eq:apdx_epf_kappa}
\end{align}

\noindent By definition, the contribution of extinct lineages to both expectations is nil: both $w_n^\beta$ and $w(t)^\beta$ vanish after extinction as we assumed that $\beta > 0$. We extend the definitions to $\beta = 0$ by taking the limit as $\beta \rightarrow 0$. For $\beta = 1$ we recover the original log partition functions in Eqs.~\eqref{eq:pop_kappa_N} and \eqref{eq:pop_kappa_T}. 

The extended functions \eqref{eq:apdx_epf_kappa} allow us to consider the statistics of forward lineages ($\beta = 0$) and ancestral lineages ($\beta = 1$). 
As shown in Eq.~\eqref{eq:apdx_ld_duality_ext}, the extended log partition functions for the two ensembles are related by an extension of our main formula \eqref{eq:pop_duality}:
\begin{align}
    \tilde \kappa_T(\tilde \kappa_N(\alpha, \beta), \beta) &= \alpha.
\end{align}

\noindent Thus thermodynamic duality holds for any fixed value of $\beta$. Both $\tilde \kappa_N$ and $\tilde \kappa_T$ are convex functions, decreasing in the first argument and increasing in the second. Note that $\tilde \kappa_N(0, 0) \leq 0$ and $\tilde \kappa_T(0, 0) \leq 0$. The two will be negative if lineages can become extinct. 

Differentiating the extended log partition functions at the point $\tilde \kappa_T(0, 1) = \Lambda$ yields
\begin{align}
    \partial_2 \tilde \kappa_N(\Lambda, 1) &= \lim_{n\rightarrow \infty} \E_b\left[\frac{\log w_n}{n} \right] := \E_b[ \log \mu ], \label{eq:pop_tilde_kappa_N_prime_bwd}
\end{align}

\noindent where we define $\log \mu$ as the average log multiplicity per generation of the ancestral process. Alternatively we can compute
\begin{align}
    \partial_2 \tilde \kappa_T(0, 1) &= \lim_{t\rightarrow \infty} \E_b\left[\frac{\log w(t)}{t} \right]. \label{eq:pop_tilde_kappa_T_prime_bwd}
\end{align}

\noindent Differentiating \eqref{eq:apdx_ld_duality_ext} with respect to $\beta$ implies
\begin{align}
    \lim_{t\rightarrow \infty} \E_b\left[\frac{\log w(t)}{t} \right] &= \frac{\E_b[ \log \mu ]}{\E_b[\tau]},
\end{align}

\noindent which is another form of the law of large numbers. A convexity argument similar to that in Eq.~\eqref{eq:pop_bwd_el_ineq} yields
\begin{align}
    0 &\geq \tilde \kappa_T(0, 0) \geq \tilde \kappa_T(0, 1) - \partial_2 \tilde \kappa_T(0, 1) = \Lambda - \frac{\E_b[\log \mu]}{\E_b[\tau]},
\end{align}

\noindent which implies
\begin{align}
    \Lambda &\leq \frac{\E_b[\log \mu]}{\E_b[\tau]}. \label{eq:apdx_kappa_ext_bregman}
\end{align}

\noindent This inequality is not equivalent to Eq.~\eqref{eq:pop_bwd_el_ineq}, as shown in \ref{apdx:counterexample}.

Much like the point $\tilde \kappa_T(0, 1) = \Lambda$ encodes statistics of ancestral lineages in a population, we can define various other points on the graph that correspond to different, but related distributions. If lineages survive indefinitely, the point $\tilde \kappa_N(0, 0) = 0$ encodes the asymptotic forward distribution. Such a distribution cannot be uniquely defined where lineages can die out; the asymptotic backward distribution is always defined, since ancestral lineages cannot go extinct by definition. If the asymptotic forward distribution is defined, it satisfies the identities
\begin{align}
    -\partial_1 \tilde \kappa_N(0, 0) &= \lim_{n\rightarrow \infty} \E_f\left[\frac{t_n}{n} \right] := \E_f[\tau], \label{eq:pop_tilde_kappa_N_prime_0} \\
    -\partial_1 \tilde \kappa_T(0, 0) &= \lim_{t\rightarrow \infty} \E_f\left[\frac{n(t)}{t} \right], \label{eq:pop_tilde_kappa_T_prime_0}
\end{align}

\noindent and from \eqref{eq:apdx_ld_duality_ext} we recover the law of large numbers for forward lineages,
\begin{align}
    \lim_{t\rightarrow \infty} \E_f\left[\frac{n(t)}{t} \right] &= \frac 1 {\E_f[\tau]}. \label{eq:pop_lln_fwd}
\end{align}

\noindent A convexity argument mirroring \eqref{eq:apdx_kappa_ext_bregman} shows that
\begin{align}
    \Lambda &= \tilde \kappa_T(0, 1) \geq \tilde \kappa_T(0, 0) + \partial_2 \tilde \kappa_T(0, 0) = \frac{\E_f[\log \mu]}{\E_f[\tau]},
\end{align}

\noindent which is equivalent to
\begin{align}
    \Lambda &\geq \frac{\E_f[\log \mu]}{\E_f[\tau]}. \label{eq:apdx_fwd_bregman}
\end{align}

\noindent A special case of this inequality was previously derived in \cite{yamauchi_unified_2022}. In contrast to Eq.~\eqref{eq:pop_bwd_el_ineq}, this inequality with the numerator replaced by $\log \E_f[\mu]$ does not always hold for variable offspring numbers as we show in \ref{apdx:counterexample}.

\section{Selection coefficients}

\label{apdx:sensitivity}

To derive Eq.~\eqref{eq:selection_kappa}, we formally consider a perturbation of the forward lineage that affects the growth rate as in Eq.~\eqref{eq:perturb_lambda} and the function $\kappa_N$ as
\begin{align}
    \kappa_N(\alpha) \mapsto \kappa_N(\alpha) + (\delta \kappa_N)(\alpha).
\end{align}

\noindent Using Eq.~\eqref{eq:pop_R0_Lambda_quad_T} for the perturbed population yields
\begin{align}
    0 &= \kappa_N(\Lambda + \delta \Lambda) + (\delta \kappa_N)(\Lambda + \delta \Lambda) \approx \kappa_N(\Lambda) + (\delta \Lambda) \kappa'_N(\Lambda) + (\delta \kappa_N)(\Lambda),
\end{align}

\noindent neglecting higher-order terms. But our Euler-Lotka equation \eqref{eq:pop_R0_Lambda_quad_T} implies that the first summand vanishes, and using Eq.~\eqref{eq:pop_kappa_prime_bwd_T} we get
\begin{align}
    (\delta \Lambda) \E_b[\tau] &= (\delta \kappa_N)(\Lambda).
\end{align}

\noindent We now show that the left-hand side is approximately the selection coefficient $s$. Following \cite{ilker_modeling_2019}, the selection coefficient is defined as
\begin{align}
    N_{\textrm{mut}}(n \overline{\tau}) &= N_{\textrm{wt}}(n \overline{\tau}) (1 + s)^n, \label{eq:s_definition}
\end{align}

\noindent where \textit{wt} and \textit{mut} denote the wild-type and mutant populations, respectively, and $\overline{\tau}$ is the typical generation length in the wild-type population, which is given by $\E_b[\tau]$. Plugging Eq.~\eqref{eq:first} into Eq.~\eqref{eq:s_definition} yields
\begin{align}
    e^{n (\Lambda + \delta \Lambda) \overline{\tau}} &\approx e^{n\Lambda \overline{\tau}} (1 + s)^n,
\end{align}

\noindent which suggests the formula
\begin{align}
    s &= e^{(\delta \Lambda) \overline{\tau}} - 1 \approx (\delta \Lambda) \E_b[\tau], \label{eq:s_formula}
\end{align}

\noindent as claimed.



If we assume that the perturbation is of the form given in Eqs.~\eqref{eq:perturb_m} and \eqref{eq:perturb_tau}, then to first order we can write
\begin{align}
    \E_f\left[ \prod_{i=1}^n (m_i + \delta m_i) \, e^{-(\Lambda + \delta \Lambda) (\tau_i + \delta \tau_i)} \right] &\approx \E_f[ w_n e^{-\Lambda t_n}] + \sum_{i=1}^n \E_f\left[ w_n e^{-\Lambda t_n} \left(\frac{\delta m_i}{m_i}\right)\right] - \Lambda \sum_{i=1}^n \E_f[ w_n e^{-\Lambda t_n} (\delta \tau_i) ] \\
    &- (\delta \Lambda) \sum_{i=1}^n \E_f[w_n e^{-\Lambda t_n} \tau_i ]. \nonumber
\end{align}

\noindent Taking logarithms and dividing by $n$ yields
\begin{align}
    \frac 1 n \log \E_f \left[ \prod_{i=1}^n (m_i + \delta m_i) \, e^{-(\Lambda + \delta \Lambda) (\tau_i + \delta \tau_i)} \right] &\approx \frac 1 n \log \E_f[ w_n e^{-\Lambda t_n} ] + \frac 1 n \sum_{i=1}^n \frac{\E_f\left[ w_n e^{-\Lambda t_n} \left(\frac{\delta m_i}{m_i}\right)\right]}{\E_f\left[ w_n e^{-\Lambda t_n}\right]} \\
    &\qquad - \Lambda \frac 1 n \sum_{i=1}^n \frac{\E_f[ w_n e^{-\Lambda t_n} (\delta \tau_i) ]}{\E_f\left[ w_n e^{-\Lambda t_n}\right]} - (\delta \Lambda) \frac 1 n \sum_{i=1}^n \frac{\E_f[w_n e^{-\Lambda t_n} \tau_i ]}{\E_f\left[ w_n e^{-\Lambda t_n}\right]}.\nonumber
\end{align}

\noindent Taking the limit as $n \rightarrow \infty$, we can interpret each of the three fractions as expectations over the backward lineage distribution $p_b$ (for the original population), see \ref{apdx:kappa_int}. Since the original and the perturbed population satisfy Eq.~\eqref{eq:pop_el_implicit}, we therefore obtain the relation
\begin{align}
    0 &\approx \E_b\left[\frac{\delta m}{m}\right] - \Lambda \E_b[\delta \tau ] - (\delta \Lambda) \E_b[\tau], \label{eq:sensitivity_almost}
\end{align}

\noindent which together with Eq.~\eqref{eq:s_formula} yields Eq.~\eqref{eq:selection_coefficient}.

\section{Perron-Frobenius theory}

\label{apdx:perron}

In this section we review the basic elements of Perron-Frobenius theory as discussed in \cite{bapat_nonnegative_1997}. Call a nonnegative matrix $M \geq 0$ \emph{primitive} if $M^n$ has all positive entries for some $n \geq 1$. The Perron-Frobenius theorem states that a primitive matrix $M$ has a unique positive eigenvalue $\lambda > 0$ of multiplicity $1$ such that all other eigenvalues have norm strictly less than $\lambda$. In particular, $\lambda$ coincides with the spectral radius $\rho(M)$. Furthermore, the left and right eigenvectors corresponding to $\lambda$ have positive entries.

For a primitive matrix $M$ with dominant eigenvalue $\lambda$ the system
\begin{align}
    (\lambda - M) x = b \label{eq:apdx_pf_fredholm_eq}
\end{align}

\noindent with $b \geq 0$ has no solution unless $b = 0$. Indeed, by the Fredholm alternative, Eq.~\eqref{eq:apdx_pf_fredholm_eq} has a solution if and only if
\begin{align}
    b \perp \ker(\lambda - M^T).
\end{align}

\noindent The kernel is spanned by the dominant left eigenvector $v$ of $M$, which has positive entries. Since $b \geq 0$ only has nonnegative entries, the two cannot be orthogonal unless $b = 0$.

\section{Multitype branching processes}

\subsection{Long-term behavior}
\label{apdx:branch}

The formula for the next-generation matrix Eq.~\eqref{eq:appl_ngm} follows from the law of total probability. The number of offspring $m$ of an individual of type $x$ is given by $p(m \given x)$, and conditional on $m$, the offspring $y$ and birth time $\tau$ are jointly distributed according to $K_m(y, \tau \given x)$. We obtain Eq.~\eqref{eq:appl_ngm} by marginalizing out $\tau$ and summing over $m \geq 1$.

Now assume we introduce a death rate $\alpha > 0$, so that the parent organism dies at a random time $t_d \sim \mathrm{Exp}(\alpha)$. In the scenario above, this prevents any offspring after time $t_d$ from being born. Conditioned on $m$, the average number of offspring of type $y$ is therefore given by
\begin{align}
    m \int_0^\infty K_m(y, \tau \given x) \, p(t_d > \tau) \dif \tau.
\end{align}

\noindent Eq.~\eqref{eq:appl_M_tilt} follows directly by plugging in the distribution for $t_d$.

Let $m_n(y,x)$ be the expected number of offspring of type $y$ produced by an individual of type $x$ after exactly $n$ generations. By the law of total expectation we can then write
\begin{align}
    m_{n+1}(y, x) &= \sum_z m_{n}(y, z) m_1(z, x).
\end{align}

\noindent Since $m_1(z, x)$ is given by the next generation matrix, we find inductively that $m_n = M^n$.

To obtain the asymptotic population distribution for a multitype age-dependent branching process, let $N(t, y)$ be the number of individuals of type $y$ born at time $t$ in a large population. Then we have the recurrence relation
\begin{align}
    N(t, y) &= \int_0^t \sum_x N(t - \tau, x) \sum_{m \geq 1} p(m \given x) \, m\, K_m(y, \tau \given x) \dif \tau. \label{eq:apdx_branch_popdist_rec}
\end{align}

\noindent In the steady growth phase where $N(t) \approx N_0 e^{\Lambda t}$, the distribution of newborn individuals converges to the population distribution, that is,
\begin{align}
    N(t, y) = N(t)\, \pi_p(y).
\end{align}

\noindent Plugging this into Eq.~\eqref{eq:apdx_branch_popdist_rec} yields Eq.~\eqref{eq:appl_dist_tree}. 

An important, but subtle question is whether the expected population size matches the behavior of typical populations, ie.~if
\begin{align}
    \E[N(t)] &\sim e^{\Lambda t} \label{eq:apdx_kesten_mean}
\end{align}

\noindent implies that
\begin{align}
    N(t) &\sim e^{\Lambda t} \label{eq:apdx_kesten_as}
\end{align}

\noindent for every population, assuming nonextinction. For general branching processes this is guaranteed by the Kesten-Stigum theorem \cite{athreya_branching_1972}, which states that under suitable regularity conditions,
\begin{align}
    \lim_{t \rightarrow \infty} N(t) \, e^{-\Lambda t} &= Z
\end{align}

\noindent almost surely and in mean, where $Z$ is a random variable with finite, positive mean. More generally, our approach is valid as long as the asymptotic behaviour of the population matches the behaviour of a typical lineage as in Eq.~\eqref{eq:pop_asymp}, which is the case if fluctuations in lineages are not too large.


\subsection{Backward process}

\label{apdx:branch_bwd}

Here we compute the asymptotic backward process of an individual, which is itself a Markov renewal process, in two different ways using Eqs.~\eqref{eq:pop_bwd_N} and \eqref{eq:pop_bwd_T}. Start with the $t$-ensemble and fix an organism $x$. Following Eq.~\eqref{eq:pop_bwd_T}, the probability of producing an offspring of type $y$ at time $\tau$ in the ancestral distribution for fixed time $t \gg \tau$ is proportional to its forward probability multiplied by the expected fitness of the offspring at time $t$, given by
\begin{align}
    a_T(x, t) &= \E[ N(t) \given x ] = \E_f[ w(t) \given x ].
\end{align}

\noindent The transition kernel of the backward distribution is then defined by
\begin{align}
    p_b(y, \tau \given x) &\propto \sum_{m \geq 1} p(m \given x) \, m\, K_m(y, \tau \given x) \, a_T(y, t - \tau). \label{eq:apdx_bwd_T_basic}
\end{align}

\noindent By summing over all $y$ and integrating over $\tau$, we find that the proportionality constant equals
\begin{align}
    a_T(x, t) &= \sum_y \sum_{m \geq 1} p(m \given x) \, m \int_0^t K_m(y, \tau \given x) \, a_T(y, t - \tau) \dif \tau.  \label{eq:apdx_bwd_NT_recur}
\end{align}

\noindent For large $t$ we can use the asymptotics
\begin{align}
    a_T(x, t) &\approx a_T(x) e^{\Lambda t}, \label{eq:apdx_bwd_aT}
\end{align}

\noindent where $a_T(x)$ can be interpreted as the reproductive value of type $x$ in the $t$-picture. Plugging this equation into Eq.~\ref{eq:apdx_bwd_NT_recur} shows that $a_T(x)$ is a left eigenvector of $M_{(-\Lambda)}$ with eigenvalue $1$. Furthermore, using Eq.~\eqref{eq:apdx_bwd_T_basic} and letting $t \rightarrow \infty$ we obtain
\begin{align}
    p_b(y, \tau \given x) &= \sum_{m \geq 1} p(m \given x) \, m\, K_m(y, \tau \given x) \, e^{- \Lambda \tau} \frac{a_T(y)}{a_T(x)}. \label{eq:apdx_bwd_T}
\end{align}

\noindent This describes the ancestral process in the $t$-picture as a Markov renewal process. Integrating over the dwelling time $\tau$ we obtain the transition kernel in Eq.~\eqref{eq:appl_Kb}.

We can derive the same ancestral distribution using Eq.~\eqref{eq:pop_bwd_N} in the $n$-ensemble. For this we define the expected fitness of a lineage in the $n$-th generation as
\begin{align}
    a_N(x, n) &= \E_f[ w_n \, e^{-\Lambda t_n} \given x ],
\end{align}

\noindent and compute
\begin{align}
    p_b(y, \tau \given x) &\propto \sum_{m \geq 1} p(m \given x) \, m\, K_m(y, \tau \given x) \, e^{-\Lambda \tau} a_N(y, n-1). \label{eq:apdx_bwd_N_basic}
\end{align}

\noindent The analogue to Eq.~\eqref{eq:apdx_bwd_NT_recur} becomes the identity
\begin{align}
    a_N(x, n) &= \sum_y a_N(y, n-1) M_{(-\Lambda)}(y, x).
\end{align}

\noindent As a result, for large $n$ we have
\begin{align}
    a_N(x, n) &\rightarrow a_N(x),
\end{align}

\noindent where $a_N(x)$ is a left eigenvector of $M_{(-\Lambda)}$ with eigenvalue $1$. Since we assume that $M_{(-\Lambda)}$ is primitive, this must equal $a_T(x)$. Letting $n \rightarrow \infty$ in Eq.~\eqref{eq:apdx_bwd_N_basic} then yields \eqref{eq:apdx_bwd_T_basic} again, showing that the two notions in Eqs.~\eqref{eq:pop_bwd_T} and \eqref{eq:pop_bwd_N} agree in the limit. 

\subsection{The partition function in the $t$-ensemble}

\label{apdx:kappa_T}

In this section we directly derive the extended log partition function $\tilde \kappa_T(\xi, \beta)$ adapting standard techniques found e.g.~in \cite{athreya_branching_1972}. The idea is to evaluate the function 
\begin{align}
    \tilde \kappa_T(\xi, \beta) &= \lim_{t \rightarrow \infty} \frac 1 t \log \E\left[w(t)^\beta e^{- \xi n(t)} \right],
\end{align}

\noindent by computing the expectation for finite $t$ and obtaining its asymptotic behavior using the final value theorem for Laplace transforms. 

Fixing $\xi$ and $\beta > 0$ we introduce the pathwise function
\begin{align}
    f(t; \ell) &= w(t)^\beta e^{- \xi n(t)},
\end{align}

\noindent where $\ell = (x_0, x_1, \ldots)$ is an arbitrary lineage. If the lineage survives to at least $x_1$ we have the following recursion relation:
\begin{align}
    f(t; \ell) &= \begin{cases}
        1, & 0 \leq t < t_1, \\
        m_1^\beta, e^{-\xi} f(t - t_1, \ell') & t_1 \leq t,
    \end{cases} \label{eq:apdx_part_rr_T}
\end{align}

\noindent where $\ell' = (x_1, x_2, \ldots)$ is the original lineage with the first organism removed. If on the other hand $x_0$ has no offspring but goes extinct at time $t_d$ we have
\begin{align}
    f(t; \ell) &= \begin{cases}
        1, & 0 \leq t < t_d, \\
        0, & t \geq t_d.
    \end{cases} \label{eq:apdx_part_rr_T_ext}
\end{align}

\noindent Here we assume that the lifetime of an individual of type $x$ conditioned on extinction is described by the probability distribution $K_0(\tau \given x)$. As discussed in \ref{apdx:kappa_ext}, the exact form of $K_0$ does not matter as long as its tails decay sufficiently quickly, as measured by its cumulant generating function.

To analyze the long-term behavior of $f$ we take Laplace transforms in the time variable, which converts Eq.~\eqref{eq:apdx_part_rr_T} into
\begin{align}
    \hat f(\lambda; \ell) &= \frac{1 - e^{-\lambda t_1}} {\lambda} + m_1^\beta e^{-\xi} e^{-\lambda t_1} \hat f(\lambda; \ell'), \label{eq:apdx_part_rr_lap_T}
\end{align}

\noindent and Eq.~\eqref{eq:apdx_part_rr_T_ext} into
\begin{align}
    \hat f(\lambda; \ell) &= \frac{1 - e^{-\lambda t_d}} {\lambda}. \label{eq:apdx_part_rr_lap_T_ext}
\end{align}

\noindent We remark for later that the first term on the right-hand side of Eqs.~\eqref{eq:apdx_part_rr_lap_T} and \eqref{eq:apdx_part_rr_lap_T_ext} is always strictly positive and has finite expectation for $\lambda > 0$. 

Denote the expectation of $f$ over all lineages starting with an individual of type $x$ by
\begin{align}
    \phi_t(x) &= \E[ m(t)^\beta e^{- \xi n(t)} \given x_0 = x ],
\end{align}

\noindent and let $\hat \phi_\lambda(x)$ be its Laplace transform in the $t$-variable. We obtain a recurrence relation for $\hat \phi$ by averaging \eqref{eq:apdx_part_rr_lap_T} over all lineages, which requires summing over all possible values of $x_1$, intergeneration times $\tau_1$ and multiplicities $m_1 \geq 1$:
\begin{align}
    \hat \phi_\lambda(x) &= \frac 1 {\lambda} - \frac 1 \lambda p(m = 0\given x) \, \int_0^\infty K_0(\tau \given x) \, e^{-\lambda \tau} \dif \tau - \frac 1 \lambda \sum_y \sum_{m \geq 1} p(m \given x) \, \int_0^\infty K_m(y, \tau \given x)  \, e^{-\lambda \tau} \dif \tau \\
    &\qquad + e^{-\xi} \sum_{m \geq 1} \sum_y m^\beta p_m(m \given x) \hat \phi_y(\lambda) \int_0^\infty K_m(y, \tau \given x) \, e^{-\lambda \tau} \dif \tau \nonumber \\
    &:= R_\lambda(x)+ e^{-\xi} \sum_y M_{(-\lambda,\beta)}(y, x) \, \hat \phi_y(\lambda),\label{eq:apdx_kappa_T_rec}
\end{align}

\noindent where $R_\lambda$ is an analytical function of $\lambda$ that is finite and positive for all $\lambda > 0$.

In vector notation, Eq.~\eqref{eq:apdx_kappa_T_rec} can be compactly written as
\begin{align}
    \left(1 - e^{-\xi} M_{(-\lambda,\beta)}\right) \, \hat \phi_\lambda &= R_\lambda. \label{eq:apdx_part_lap_T}
\end{align}

\noindent If the left-hand side of Eq.~\eqref{eq:apdx_part_lap_T} is invertible this is equivalent to
\begin{align}
    \hat \phi_\lambda &= \left(1 - e^{-\xi} M_{(-\lambda,\beta)}\right)^{-1} R_\lambda. \label{eq:apdx_part_lap_T_inv}
\end{align}

We know that $\phi_t(x)$ grows asymptotically as
\begin{align}
    \phi_t(x) &\approx e^{\kappa_T(\xi,\beta) t}. \label{eq:apdx_part_lap_T_asymp}
\end{align}

\noindent Eq.~\eqref{eq:apdx_part_lap_T_asymp} implies that the Laplace transform of $\phi$ is finite whenever $\lambda > \kappa_T(\xi,\beta)$, and that it blows up at $\lambda = \kappa_T(\xi,\beta)$. We can therefore compute $\kappa_T(\xi,\beta)$ by finding the largest real pole of $\hat \phi$.

For large enough $\lambda > 0$, $R_\lambda$ is finite and positive and the inverse in Eq.~\eqref{eq:apdx_part_lap_T_inv} exists. As we decrease $\lambda$, either of the two terms on the right-hand side can blow up. The inverse blows up when
\begin{align}
    \spr(e^{-\xi} M_{(-\lambda,\beta)}) &= 1, \label{eq:apdx_part_spr_bd}
\end{align}

\noindent in which case Eq.~\eqref{eq:apdx_part_lap_T} has no solution by the positivity of $M_{(-\lambda,\beta)}$ and $R_\lambda$, see \ref{apdx:perron}. Thus the Laplace transform is undefined at that value of $\lambda$ and we obtain
\begin{align}
    \spr(e^{-\xi} M_{(-\kappa_T(\xi,\beta),\beta)}) &= 1,
\end{align}

\noindent which is equivalent to
\begin{align}
    \log \spr(M_{(-\kappa_T(\xi,\beta),\beta)}) &= \xi.\label{eq:apdx_branch_kappa_T}
\end{align}

\noindent Comparing this with Eq.~\eqref{eq:appl_kappa_N} yields shows that Eq.~\eqref{eq:pop_duality} holds in this case.

The other possibility is that $R_\lambda$ blows up first, which happens if the cumulant generating function of $K_0(\tau \given x)$ is not defined at $\lambda$. This can be avoided if the distribution $K_0(\tau \given x)$ decays sufficiently quickly, as we discuss in \ref{apdx:kappa_ext}. For $\beta = 1$ in a growing population, Eq.~\eqref{eq:apdx_branch_kappa_T} happens for some $\lambda > 0$, while $R_\lambda$ is always finite for $\lambda > 0$, so this scenario does not occur for a growing population.

\section{Model details}

\label{apdx:models}

\subsection{Epidemic model}

\label{apdx:epi}
The next generation matrix for the random infectivity model is given by the integral operator
\begin{align}
    M(\iota', \iota) &= \frac 1 {\sqrt{2 \pi (1 - c_\iota^2) \sigma_\iota^2}} e^{-\frac{(\iota' - c_\iota \iota - (1 - c_\iota) \overline{\iota})^2}{2 (1 - c_\iota^2) \sigma_\iota^2}} e^\iota.\label{eq:apdx_epi_M}
\end{align}

\noindent To compute its dominant eigenvalue $R_0$ we make the following Ansatz \cite{lin_effects_2017}, where $x$ is to be determined:
\begin{align}
    R_0 e^{-\frac{(\iota' - x)^2}{2 \sigma_\iota^2}} &= \int M(\iota', \iota) \, e^{-\frac{(\iota - x)^2}{2 \sigma_\iota^2}} \dif \iota = e^{x + \frac{\sigma_\iota^2}{2}} \int \frac 1 {\sqrt{2 \pi (1 - c_\iota^2) \sigma_\iota^2}} e^{-\frac{(\iota' - c_\iota \iota - (1 - c_\iota) \overline{\iota})^2}{2 (1 - c_\iota^2) \sigma_\iota^2}} e^{-\frac{(\iota - x - \sigma_\iota^2)^2}{2 \sigma_\iota^2}} \dif \iota.  \label{eq:apdx_epi_iotap_eq}
\end{align}

\noindent Up to a normalisation constant of $\sqrt{2 \pi \sigma_\iota^2}$, we can interpret the integral as the marginal distribution of $\iota'$ where
\begin{align}
    \iota &\sim \mathcal N(x + \sigma_\iota^2, \sigma_\iota), \\
    \iota' \given \iota &\sim \mathcal N\left(c_\iota \iota + (1 - c_\iota) \overline{\iota}, (1 - c_\iota^2) \sigma_\iota^2\right).
\end{align}

\noindent Comparing the means on both sides of Eq.~\eqref{eq:apdx_epi_iotap_eq} yields the compatibility condition
\begin{align}
    x &= (1 - c_\iota) \overline \iota + c_\iota x + c_\iota \sigma_\iota^2.
\end{align}

\noindent From this we obtain Eq.~\eqref{eq:epi_r0}. The dominant right eigenvector of $M$, which represents the population distribution, equals
\begin{align}
    \pi_p(\iota) &= \mathcal N\left(\iota; \overline{\iota} + \frac {c_\iota}{1 - c_\iota} \sigma_\iota^2, \sigma_\iota\right).
\end{align}

\noindent Plugging this into Eq.~\eqref{eq:apdx_epi_iotap_eq} yields Eq.~\eqref{eq:epi_r0}. Here we explicitly assume that $c_\iota < 1$ and ignore the case of perfect heritability. 

For the dominant left eigenvector of $M$ we make the following Ansatz, where $y$ is unknown:
\begin{align}
    \int e^{y \iota'} M(\iota', \iota) \dif \iota' &= R_0 e^{y \iota}.
\end{align}

\noindent Interpreting this in terms of the conditional expectation of $e^{y\iota'}$ given $\iota$, we can verify that this holds for $y = (1 - c_\iota)^{-1}$. Plugging this into Eq.~\eqref{eq:appl_Kb} yields the transition matrix
\begin{align}
    R_0^{-1} e^{\frac 1 {1 - c_\iota} \iota'} M(\iota', \iota) \, e^{-\frac 1 {1 - c_\iota} \iota} &= \frac 1 {R_0 \sqrt{2 \pi (1 - c_\iota^2) \sigma_\iota^2}}  e^{-\frac{\left(\iota' - c_\iota \iota - (1 - c_\iota) \overline \iota - (1 + c_\iota) \sigma_\iota^2\right)^2}{2 (1 - c_\iota^2) \sigma_\iota^2}},
\end{align}

\noindent which shows that the ancestral distributions follow the autoregressive process defined in Eq.~\eqref{eq:epi_iotap_a}. The asymptotic backward distribution over infectivities equals
\begin{align}
    \pi_b(\iota) &= \mathcal N\left(\iota; \overline{\iota} + \frac {1 + c_\iota}{1 - c_\iota} \sigma_\iota^2, \sigma_\iota\right).
\end{align}

Now assume that each strain has a different latent period $\tau$ inherited according to Eq.~\eqref{eq:epi_lambda_f}. The tilted next generation matrix becomes
\begin{align}
    M_{(-\alpha)}(\iota', \tau'; \iota, \tau) &= K_i(\iota', \iota) \, e^{\iota} \, K_l(\tau', \tau) \, e^{-\alpha \tau},
\end{align}

\noindent where $K_i$ and $K_l$ are the transition matrices for the infectivities and the latency period, respectively. This is the tensor product of two matrices, one involving the infectivities $\iota'$ and $\iota$ and the other the latencies $\tau'$ and $\tau$. For the spectral radius we obtain
\begin{align}
    \spr(M_{(-\alpha)}) &= R_0 \, \spr(T_{(-\alpha)}),
\end{align}

\noindent with $R_0$ still given by Eq.~\eqref{eq:epi_r0} and
\begin{align}
    T_{(-\alpha)}(\tau', \tau) &= \frac 1 {\sqrt{2 \pi (1 - c_\tau^2) \sigma_\tau^2}} e^{-\frac{(\tau' - c_\tau \tau - (1 - c_\tau) \overline{\tau})^2}{2 (1 - c_\tau^2) \sigma_\tau^2}} e^{-\alpha \tau}.
\end{align}

\noindent This is of the same form as the matrix $M$ earlier, and we can compute 
\begin{align}
    \log \spr(T_{(-\alpha)}) &= \frac {1 + c_\tau}{1 - c_\tau} \frac{\sigma_\tau^2}{2} \alpha^2 - \overline{\tau} \alpha.
\end{align}

\noindent The Euler-Lotka equation therefore reads
\begin{align}
    \frac {1 + c_\tau}{1 - c_\tau} \frac{\sigma_\tau^2}{2} \Lambda^2 - \overline{\tau} \Lambda + \log R_0 &= 0,
\end{align}

\noindent which has solution given by Eq.~\eqref{eq:epi_el}.

Since the tilted matrix $M_{(-\Lambda)}$ factors as a tensor product, infectivities $\iota$ and latency periods $\tau$ evolve independently in backward lineages. The population distribution over $\tau$ is given by the right eigenvector of $M_{(-\Lambda)}$,
\begin{align}
    \pi_p(\tau) &= \mathcal N\left(\tau; \overline{\tau} - \Lambda \frac{c_\tau}{1 - c_\tau} \sigma_\tau^2\right).
\end{align}

\noindent Similarly, the left eigenvector is given by
\begin{align}
    a(\tau) &\propto e^{-\frac{\Lambda}{1 - c_\tau}}.
\end{align}

\noindent We therefore obtain the following transition matrix for the ancestral distribution:
\begin{align}
    e^{-\frac {\Lambda} {1 - c_\tau} \tau'} M(\tau', \tau) \, e^{\frac {\Lambda} {1 - c_\tau} \tau} &\propto \frac 1 {\sqrt{2 \pi (1 - c_\tau^2) \sigma_\tau^2}}  e^{-\frac{\left(\tau' - c_\tau \tau - (1 - c_\tau) \overline \tau + \Lambda (1 + c_\tau) \sigma_\tau^2\right)^2}{2 (1 - c_\tau^2) \sigma_\tau^2}},
\end{align}

\noindent with stationary distribution
\begin{align}
    \pi_b(\tau) &= \mathcal N\left(\tau; \overline{\tau} - \Lambda \frac {1 + c_\tau}{1 - c_\tau} \sigma_\tau^2, \sigma_\tau\right).
\end{align}

In Fig.~\ref{fig:appl} Awe compare these theoretical predictions with numerical simulations. Since simulating an exponentially growing population for long times is infeasible, we instead simulate a set of $N = 1000$ lineages in parallel. Whenever a reproduction event results in $N^* > N$ lineages, we continue the simulation after sampling $N$ of these lineages at random as in a Moran process. We then compute the backward distribution by tracing the ancestry of a single random individual in the population. Model parameters were $\overline{\iota} = 0$, $\sigma_\iota=0.4$ and $c_\iota = 0.5$ for the infectivities, and $\overline{\tau} = 1$, $\sigma_\tau = 0.25$ and $c_\tau = 0.8$ for the latencies.







\subsection{Squirrel model}

\label{apdx:squirrels}

For the population model in Fig.~\ref{fig:appl}B we assume that each individual has independent generation lengths $\tau \sim \Gamma(k, \beta)$ (parametrizing the Gamma distribution by its rate $\beta$) and produces offspring with distribution $m \sim 1 + \Geom(\mu)$ (parametrizing the geometric distribution by its mean). We treat each individual as an offspring of itself since squirrels can reproduce in more than one season. 

The partition function $\kappa_N$ can be written as 
\begin{align}
    \kappa_N(\alpha) &= \lim_{n \rightarrow \infty} \log \E_f[ w_n \, e^{-\alpha t_n}] = \lim_{n \rightarrow \infty} \log \prod_{i=1}^n \E[ m_i \, e^{-\alpha \tau_i}] = \lim_{n \rightarrow \infty}\frac 1 n \log \left[ (1 + \mu) \left(\frac{\beta}{\beta + \Lambda}\right)^k \right]^n \\
    &= \log(1 + \mu) + k \log \beta - k \log (\beta + \Lambda) \nonumber,
\end{align}

\noindent since $m_i$ and $\tau_i$ are independent of each other. In Fig.~\ref{fig:appl}B we estimate $\delta \kappa_N$ in Eq.~\eqref{eq:selection_kappa} by taking derivatives of this with respect to the parameters $k$, $\beta$ and $\mu$. The growth rate $\Lambda$ for the wild-type population is given by the classical Euler-Lotka equation:
\begin{align}
    1 &= \E[ m \, e^{-\Lambda \tau}] = (1 + \mu) \left(\frac{\beta}{\beta + \Lambda}\right)^\alpha, \label{eq:el_squirrel}
\end{align}

\noindent which has the solution
\begin{align}
    \Lambda &= \beta \left((1 + \mu)^{1/\alpha} - 1\right).\label{eq:lambda_squirrel}
\end{align}

\noindent As with the epidemic example, we simulated this model while maintaining a fixed population size $N = 100$, until either the mutant of the wild-type species reaches fixation. We then estimate the fixation probability by simulating the system repeatedly and counting the number of cases where the wild-type species succeeds.

\begin{table}[]
    \centering
    \begin{tabular}{c|c|c}
        mutant & $\tau$ distribution & $m$ distribution \\ 
        \hline
        WT & $\Gamma(36, 6)$ & $\Geom(2)$ \\
        $k$$\blacktriangle$ & $\Gamma(42, 6)$ & $\Geom(2)$ \\
        $k$\rotatebox[origin=c]{180}{$\blacktriangle$} & $\Gamma(30, 6)$ & $\Geom(2)$ \\
        $\beta$$\blacktriangle$ & $\Gamma(36, 6.5)$ & $\Geom(2)$ \\
        $\beta$\rotatebox[origin=c]{180}{$\blacktriangle$} & $\Gamma(36, 5.5)$ & $\Geom(2)$ \\
        $\mu$$\blacktriangle$ & $\Gamma(36, 6)$ & $\Geom(2.05)$ \\
        $\mu$\rotatebox[origin=c]{180}{$\blacktriangle$} & $\Gamma(36, 6)$ & $\Geom(1.95)$ 
    \end{tabular}
    \caption{Parameters for the wild-type population in \cref{fig:appl}B and six different invasive species. The Gamma distribution is parametrized by its shape and rate, the geometric distribution starting at $0$ by its mean.}
    \label{tab:squirrel_param}
\end{table}

\subsection{Bacterial model}

\label{apdx:plasmids}

The growth rate of the bacterial model with no addiction and no metabolic burden ($\beta = 0$) is 
\begin{align}
    \Lambda &= \frac{\log 2}{\tau_0}, \label{eq:lambda_plasmid}
\end{align}

\noindent since all cells divide exactly at age $\tau_0$. The presence of plasmids with no metabolic burden is unbiological and results in pathological behaviour: the number of plasmids, which evolves according to Eq.~\eqref{eq:plasmid_K} in a lineage, follows a critical branching process that either reaches $0$ or grows indefinitely; in particular, it does not have a stationary distribution. We thus restrict our attention to $\beta > 0$.

If addiction is implemented and cells with $0$ plasmids have no offspring, then in the case $\beta = 0$ the population will still grow with the rate $\Lambda$ defined in Eq.~\eqref{eq:lambda_plasmid}. Indeed, the number of plasmids in the population doubles with each generation, so that there are always plasmid-bearing cells which can reproduce. Since a cell with $k > 0$ plasmids has a probability $2^{-2k+1}$ of producing infertile offspring, selection will shift the population distribution of plasmids to increasingly larger values of $k$. In this case, the average number of plasmids per cell will grow indefinitely, and addiction becomes asymptotically irrelevant. In this case there is again no stationary distribution for the number of plasmids. If $\beta > 0$, selection due to the metabolic burden ensures that the plasmid distribution stabilises, and $\E_b[k]$ is well-defined and finite.

In Fig.~\ref{fig:appl}C, we simulated a population with addiction and with $\beta = 0.1\%$ with fixed carrying capacity $N = 1000$. The growth rate of the population was estimated as described by the cloning algorithm of \cite{lecomte_numerical_2007,giardina_direct_2006}, which is asymptotically exact as $N \rightarrow \infty$; in our case the results were visually indistinguishable for $N = 1000$ and $N = 2000$. We approximated Eq.~\eqref{eq:selection_coefficient} by sampling the ancestry of a random individual in the population as in \ref{apdx:epi} and averaging Eqs.~\eqref{eq:plasmid_delta_m} and \eqref{eq:plasmid_delta_tau} over individuals. Our predictions are slightly biased upwards due to our comparison against the case $\beta = 0$, which we showed above results in qualitatively different behaviour.

\section{A simple counterexample}

\label{apdx:counterexample}

In this section we consider a simple model that provides a counterexample to a strengthening of Eq.~\eqref{eq:apdx_fwd_bregman} and shows that neither of Eqs.~\eqref{eq:pop_bwd_el_ineq} and \eqref{eq:apdx_kappa_ext_bregman} implies the other. Consider a population consisting of two types, $A$ and $B$, where type $A$ produces two offspring at age $1$ and type $B$ produces $2N$ offspring at age $T$, where $N, T > 1$. Each offspring is uniformly sampled from $A$ and $B$, and in particular we have
\begin{align}
    R_0 &= 1 + N, & \E_f[\tau] &= \frac {T+1} 2.
\end{align}

\noindent The tilted next-generation matrix is
\begin{align}
    M_{(-\Lambda)} &= \begin{bmatrix} e^{-\Lambda} & N e^{-\Lambda T} \\ e^{-\Lambda} & N e^{-\Lambda T} \end{bmatrix}.
\end{align}

\noindent Since this is of rank $1$, the EL equation reads
\begin{align}
    e^{-\Lambda} + N e^{-\Lambda T} &= 1. \label{eq:apdx_counter_el}
\end{align}

\noindent As each organism is assigned a type at random, the forward and population distributions are both uniform over $A$ and $B$, whereas the ancestral distribution is determined by the dominant left eigenvector of $M_{(-\Lambda)}$. 

The two summands in the Euler-Lotka equation \eqref{eq:apdx_counter_el} represent the reproductive value of the two types. They are equal if and only if $e^{-\Lambda} = 1/2$, ie.
\begin{align}
    \Lambda &= \log(2), & N &= 2^{T-1}.
\end{align}

\noindent In this case the backward distribution over types equals the population distribution. For $T$ large enough,
\begin{align}
    \frac{\E_f[\log \mu]}{\E_f[\tau]} < \Lambda < \frac{\log R_0}{\E_f[\tau]},
\end{align}

\noindent which shows that Eq.~\eqref{eq:apdx_fwd_bregman} with $\E_f[\log \mu]$ replaced by $\log R_0$ does not necessarily hold for variable offspring numbers.

In general, let
\begin{align}
    r &:= \frac{N e^{-\Lambda T}}{e^{-\Lambda}} = \frac{\pi_b(B)}{\pi_b(A)}.
\end{align}

\noindent Fixing $N > 1$, in the regime of small $r$ (or equivalently, large $T$),
\begin{align}
    \log(1 + N) = \log(R_0) &\geq \E_b[\log \mu] = \log(2) + \frac{r}{1 + r} \log(N).
\end{align}

\noindent In contrast, for large $r$ (small $T$),
\begin{align}
    \log(1 + N) = \log(R_0) &\leq \E_b[\log \mu] = \log(2) + \frac{r}{1 + r} \log(N),
\end{align}

\noindent which shows that the inequalities \eqref{eq:pop_bwd_el_ineq} and \eqref{eq:apdx_kappa_ext_bregman} are not equivalent.



\end{document}